\documentclass[a4paper,fleqn,usenatbib]{mnras}

\usepackage{newtxtext,newtxmath}
\usepackage{times}

\usepackage{graphicx}	
\usepackage{amsmath}	
\usepackage{amssymb}	

\usepackage{bm} 
\usepackage{multirow} 	
\usepackage{chngcntr} 
\counterwithout*{figure}{section}
\counterwithout*{table}{section}

\title[Resolved CO (1--0) molecular gas in lensed AGN]{SHARP -- VI. Evidence for CO (1--0) molecular gas extended on kpc-scales in AGN star forming galaxies at high redshift}
\author[C. Spingola et al.]{C.~Spingola$^{1,2,3}$\thanks{E-mail:  spingola@ira.inaf.it},
J.~P.~McKean$^{1,4}$, S.~Vegetti$^{5}$, D.~Powell$^{5}$, M.~W.~Auger$^{6,7}$, 
\newauthor L.~V.~E.~Koopmans$^{1}$, C.~D.~Fassnacht$^8$, D.~J.~Lagattuta$^{9,10,11}$, F.~Rizzo$^{5}$, H.~R. Stacey$^{1,4}$, 
\newauthor and F.~Sweijen$^{12}$
\\
$^1$Kapteyn Astronomical Institute, Postbus 800, NL-9700 AV Groningen, the Netherlands\\
$^2$INAF $-$ Istituto di Radioastronomia, Via Gobetti 101, I$-$40129, Bologna, Italy \\
$^3$Dipartimento di Fisica e Astronomia, Universit\`{a} degli Studi di Bologna, Via Gobetti 93/2, I$-$40129 Bologna, Italy \\
$^4$ASTRON, Netherlands Institute for Radio Astronomy, Oude Hoogeveensedijk 4, 7991 PD Dwingeloo, the Netherlands\\
$^5$Max Planck Institute for Astrophysics, Karl-Schwarzschild-Strasse 1, 85740 Garching, Germany\\
$^6$Institute of Astronomy, University of Cambridge, Madingley Rd, Cambridge CB3 0HA, United Kingdom\\
$^7$Kavli Institute for Cosmology, University of Cambridge, Madingley Rd, Cambridge CB3 0HA, United Kingdom \\
$^8$Department of Physics, University of California, Davis, CA 95616, USA\\
$^9$Univ Lyon, Ens de Lyon, CNRS, Centre de Recherche Astrophysique de Lyon UMR5574, F-69230 Saint-Genis-Laval, France\\
$^{10}$Centre for Extragalactic Astronomy, Department of Physics, Durham University, Durham DH1 3LE, UK\\
$^{11}$Institute for Computational Cosmology, Durham University, South Road, Durham DH1 3LE, UK\\
$^{12}$Leiden Observatory, Leiden University, PO Box 9513, NL$-$2300RA Leiden, the Netherlands
}

\date{Accepted 2020 May 11. Received 2020 May 08; in original form 2019 May 06}

\pubyear{2020}

\begin{document}
\label{firstpage}
\pagerange{\pageref{firstpage}--\pageref{lastpage}}
\maketitle

\begin{abstract}
\noindent We present a study of the stellar host galaxy, CO (1--0) molecular gas distribution and AGN emission on 50 to 500 pc-scales of the gravitationally lensed dust-obscured AGN MG~J0751+2716 and JVAS~B1938+666 at redshifts 3.200 and 2.059, respectively. By correcting for the lensing distortion using a grid-based lens modelling technique, we spatially locate the different emitting regions in the source plane for the first time. Both AGN host galaxies have 300 to 500 pc-scale size and surface brightness consistent with a bulge/pseudo-bulge, and 2 kpc-scale AGN radio jets that are embedded in extended molecular gas reservoirs that are 5 to 20 kpc in size.  The CO (1--0) velocity fields show structures possibly associated with discs (elongated velocity gradients) and interacting objects (off-axis velocity components). There is evidence for a decrement in the CO (1--0) surface brightness at the location of the host galaxy, which may indicate radiative feedback from the AGN, or offset star formation.We find CO--H$_2$ conversion factors of around $\alpha_{\rm CO} = 1.5\pm0.5$~(K km\,s$^{-1}$ pc$^2$)$^{-1}$, molecular gas masses of $> 3\times10^{10}$~M$_{\odot}$, dynamical masses of $\sim 10^{11}$~M$_{\odot}$ and gas fractions of around 60 per cent. The intrinsic CO line luminosities are comparable to those of unobscured AGN and dusty star-forming galaxies at similar redshifts, but the infrared luminosities are lower, suggesting that the targets are less efficient at forming stars. Therefore, they may belong to the AGN feedback phase predicted by galaxy formation models, because they are not efficiently forming stars considering their large amount of molecular gas.

\end{abstract}




\section{Introduction}
\label{sec:intro}
The bulk of the stellar population in the Universe formed between redshift 1 and 3, when the comoving cosmic star formation rate (SFR) density peaks and the active galactic nuclei (AGN) were at their peak \citep{Hopkins2006,MadauDickinson2014}. The co-evolution of these two processes has led to the suggestion that negative feedback from AGN is responsible for the subsequent decline in star formation (e.g. \citealt{Croton2006}). However, this does not directly explain the increased star formation efficiency in individual galaxies compared to the local Universe. Indeed, there is even growing support (e.g. \citealt{Maiolino2017}) for AGN-induced positive feedback (e.g. \citealt{Ishibashi2012}), and the role of AGN can only be definitively clarified by spatially resolving the molecular gas that fuels the star formation in these systems (e.g. \citealt{Nesvadba2017,Leung2017}). 

Observations of high redshift galaxies at mm-wavelengths have shown that they have large molecular gas reservoirs ($\geq 10^{9-10}$ M$_{\odot}$); since the cold molecular gas is a fundamental ingredient of star formation, galaxies that show molecular line emission have become the most important population to study the evolution of cosmic star formation and its connection to the evolution of AGN \citep{Fabian2012,Carilli2013}.  The ground-state rotational transition  $J=1\rightarrow\,0$ of the CO molecule (rest-frame 115.271~GHz) is generally used to trace and measure the total mass of the cold molecular gas reservoir in galaxies. This is because CO is the most abundant molecule after H$_{2}$,  the latter of which does not have a permanent dipole moment and, therefore, cannot be observed directly. 

Many detections of CO emission in active galaxies at $z\sim2$--3 were made possible in the last decade thanks to the increased sensitivity and large-bandwidth of the upgraded Ka-band receivers  ($\sim26.5$ to 40~GHz) on the Green Bank Telescope (GBT) and the Karl G. Jansky Very Large Array (VLA, e.g. \citealt{Emonts2014}). Moreover, recently a significant observational effort using wide-field surveys has been devoted to blindly detecting CO (1--0) at high redshift and understanding its evolution across cosmic time \citep[e.g.][]{Aravena2016, Decarli2019, Riechers2019}.

In a few cases, the CO emission from the gas reservoirs of high redshift galaxies hosting an AGN has been detected with interferometric arrays, revealing that in these objects the cool interstellar medium (ISM) is distributed in relatively compact ($<$ few kpc) regions \citep{Carilli2002, Walter2004, Riechers2008, Riechers2009,Riechers2011}. Therefore, the inferred SFR surface densities in these galaxies are close to the Eddington limit ($\sim10^3$~M$_{\odot}$~yr$^{-1}$~kpc$^{-2}$; \citealt{Walter2009, Carilli2013}), suggesting that these galaxies are also undergoing a starburst phase together with the growth of the super massive black hole. The full width at half maximum (FWHM) of the CO molecular gas in high redshift galaxies has been observed with velocities up to $\sim 1000$ km\,s$^{-1}$, suggesting the presence of AGN-driven outflows (e.g. \citealt{Ivison2012}). The outflows are thought to be due to the AGN radio jets, which are transporting material out to several kpc distances from the central black hole. Moreover, the line profile of jet-driven outflows generally requires multiple Gaussian components and can show broad faint wings (e.g. \citealt{Feruglio2015}).

Most of the ultra-luminous quasar--starburst galaxies that have been observed so far tend to be part of interacting or merging systems. The merging/interaction process can trigger bursts of star formation and the fuelling of AGN by providing a mechanism for the gas to fragment and collapse \citep{Kennicutt1987, Sanders1988}. The merging process leads to a disturbed morphology that often results in tidal tails and a complete disruption of the nuclear and outer parts of the merging galaxies \citep{Toomre1972}. Moreover, high resolution numerical simulations suggest that merger remnants can become galaxies with a cold molecular gas disc \citep{Governato2007,Governato2009,Hopkins2009b,Hopkins2009a}. This is in agreement with observations; in some high redshift star-forming galaxies, evidence has been found for star formation occurring in discs that are potentially associated with this merging formation scenario \citep{Genzel2006,Tacconi2008,Engel2010}. In some cases, the merging process can lead to the formation of compact elliptical galaxies \citep{vanDokkum2008,vanDokkum2015}, whose surface brightness profile can be well described by two S\'ersic components, one spheroidal-like and one disc-like. This is consistent with an inside-out growth scenario, in which the central compact core is formed first and the surrounding extended stellar halo is accreted later via minor mergers \citep{Oldham2017}.

It is possible to study star-forming galaxies hosting an AGN through mapping their spectral energy distribution (SED), especially in the far-infrared (FIR) to near-infrared (NIR) wavelength range (1--1000~$\mu$m), where the AGN activity has several spectral features from the accretion disc, the hot torus and the cold dust (e.g. \citealt{Drouart2016,Podigachoski2016}). Typically, the SED of a star-forming galaxy peaks at $\sim 100~\mu$m, whereas when the dust is heated directly by the central AGN, the resulting peak in the thermal component dominates at shorter wavelengths, corresponding to a hotter effective dust temperature. Even if the SED features are distinct, the relative importance of star formation and AGN activity to the respective dust components is difficult to determine at FIR wavelengths and is still a matter of debate \citep{Wuyts2011,Hayward2015, Ciesla2015, Stacey2018}.

The difficulties in spatially resolving the cold molecular gas distribution are related to the limitations in angular resolution of interferometric arrays and the low surface brightness of the CO (1--0) emission. However, by observing galaxies that are magnified by a gravitational lens, it is possible to obtain higher spatial resolution and sensitivity observations of star-forming galaxies and AGN at cosmologically interesting epochs. When such observations are coupled with advanced gravitational lens modelling algorithms for the analysis of multi-wavelength data, it is also possible to recover the intrinsic morphology of the background object and, therefore, the intrinsic molecular gas distribution can be resolved with respect to the host galaxy (e.g. \citealt{Swinbank2011,Danielson2011,Weiss2013,Rybak2015a,Rybak2015b}). However, for the several studies of gravitational lensing systems carried out thus far, it was difficult to robustly infer the morphology of the CO (1--0) emission, because of the limited angular resolution of the observations \citep{Riechers2011,Sharon2016}. Also, without properly resolved datasets, the magnification of the molecular gas component is  unknown, even in those cases where the lens model is robust, which limits the interpretation of the observations of gravitationally lensed galaxies \citep{Deane2013}.

In this paper, we use the gain in angular resolution and sensitivity provided by strong gravitational lensing to carry out a resolved study of the CO (1--0) molecular gas properties of two radio-loud AGN on unprecedented angular scales. These data are coupled with high angular resolution continuum imaging at optical/near-infrared (NIR) and radio wavelengths to investigate the distribution and kinematics of the molecular gas relative to the stellar emission and the non-thermal jets produced from the central engine. Our aim is to better understand the build-up of the stellar population in AGN host galaxies at redshifts 2 to 3 and determine to what extent mechanical feedback processes are affecting the evolution of these galaxies. In Section \ref{sec:targets}, we introduce the two targets and in Section \ref{sec:observations} we describe the multi-wavelength observations (new and archival), and the data reduction processes. The observed image-plane properties of the two lensed AGN are presented in Section \ref{sec:results-imageplane}. In Section \ref{sec:results}, we present the lens modelling procedure that we have applied to the different observations and describe the intrinsic source-plane properties. In Section \ref{Sec:results_3} we discuss the molecular gas mass, dynamical mass and gas fraction of the two systems. The discussion of the results and our conclusions are in Sections \ref{sec:discussion} and \ref{sec:conclusions}, respectively.  

Throughout this paper, we assume $H_0=70\; \mathrm{km\,s^{-1}~Mpc^{-1}}$, $\Omega_{\rm M}=0.31$, and $\Omega_{\Lambda}=0.69$ \citep{Planck2016}. The spectral index $\alpha$ is defined as $S_{\nu} \propto \nu^{\alpha}$, where $S_{\nu}$ is the flux density as a function of frequency $\nu$.

\section{Targets}
\label{sec:targets} 

We have observed two radio-loud AGN as part of this study. They were selected due to previous detections of CO (1--0) at lower angular resolution with either the GBT and/or the VLA, due to their extensive optical/IR imaging and also because they have the largest angular extent (and most powerful) jets of all known lensed radio sources. In this section, we give a short review of our two targets.

\subsection{MG~J0751+2716} 

MG~J0751+2716 was found as part of the MIT--Green Bank survey for lensed radio sources by \citet{Lehar1997}. The background AGN is at redshift $z_s = 3.200\pm 0.001$, based on the detection of five narrow-emission lines in the rest-frame ultra-violet (UV) part of the spectrum, and the lens is a massive elliptical galaxy at redshift $z_l=0.3502\pm0.0002$ \citep{Tonry1999}. Spectroscopy of the environment surrounding the lensing galaxy has shown that it is part of a larger group of galaxies with 26 confirmed members and a velocity dispersion of $\sigma_{\rm group} = 400^{+60}_{-70}$~km\,s$^{-1}$ \citep{Lehar1997,Wilson2016}. The background radio source has a complex core-jet structure with an intrinsic projected size of 1.2~kpc, which, when gravitationally lensed, forms large arcs that have been detected on mas-scales with Very Long Baseline Interferometry (VLBI; \citealt{Spingola2018}). The AGN host galaxy has also been found to be quite bright at optical and NIR wavelengths, with evidence of a gravitational arc from high resolution imaging with the {\it Hubble Space Telescope} ({\it HST}; \citealt{Alloin2007}).

MG~J0751+2716 has extensive detections of CO molecular line emission; and CO (1--0), (4--3), (3--2) and (8--7) observations show it has a molecular gas content that is similar to quasar host galaxies at comparable redshifts \citep{Barvainis2002,Alloin2007,Riechers2011}. However, the angular resolution of previous CO (1--0) observations with the GBT and VLA could not spatially resolve the distribution of the molecular gas; the CO was found to be compact on 3~arcsec scales, with an almost Gaussian line profile, a FWHM of $350\pm70$~km\,s$^{-1}$ and a line intensity of $I_{\rm CO} = 0.550\pm0.095$~Jy~km\,s$^{-1}$ \citep{Riechers2011}.  Furthermore, dense molecular gas tracers (such as HCN and H$_2$O) were not detected in this system \citep{Carilli2005, Riechers2006water}. MG~J0751+2716 is a bright IR source, with a cold dust temperature of $T_{\rm D} = 36.2^{+1.9}_{-1.7}$~K and a dust emissivity of $\beta = 2.4\pm0.2$, both of which are consistent with star formation at the level of  $\mu_{\rm FIR}~\times~\rm{SFR} = 10^{3.9}$~M$_{\odot}$~yr$^{-1}$, where $\mu_{\rm FIR}$ is the lensing magnification factor at FIR wavelengths \citep{Stacey2018}. 

\subsection{JVAS~B1938+666}

JVAS~B1938+666 was discovered as part of the Jodrell Bank--VLA Astrometric Survey \citep{Patnaik1992, Browne1998, Wilkinson1998} by \citet{King1997}. Observations show a partial Einstein ring and double source morphology at radio wavelengths, as well as a complete Einstein ring at NIR wavelengths \citep{Rhoads1996,King1998,Tonry2000,Lagattuta2012}. The lensing galaxy is a massive elliptical at redshift $z_l = 0.881\pm0.001$ \citep{Tonry2000}. The redshift of the background AGN has been more challenging to establish, as there are no strong emission lines detected at either optical or NIR wavelengths \citep{Tonry2000,Lagattuta2012}. However, blind spectral line observations of the CO (3--2) and (2--1) transitions with the Combined Array for Research in Millimeter-wave Astronomy (CARMA)  found the source redshift to be $z_s = 2.0590\pm0.0003$ \citep{Riechers2011_b1938}. The AGN host galaxy is extremely red; it was not detected in {\it HST} imaging at optical wavelengths, but was detected in the NIR with the W. M. Keck Telescope adaptive optics \citep{Lagattuta2012}. A lensing reconstruction of the AGN host galaxy found it to have a projected size of about 1.3~kpc \citep{Vegetti2012}.

\citet{Sharon2016} detected the CO (1--0) line with the VLA on 2.5~arcsec-scales, finding that the line profile has an asymmetric double-horn structure, as is typically seen in rotating gas discs with a strong differential magnification (e.g. \citealt{Paraficz2018}). Note that this line profile was consistent with the higher order CO lines detected with CARMA. The FWHM of the CO (1--0) line is $654\pm71$~km\,s$^{-1}$ and the integrated line flux is $I_{\rm CO} = 0.93\pm0.11$~Jy~km\,s$^{-1}$. The molecular gas emission is dominated by a compact component that is coincident with the lensing system, but there is also evidence of a faint extended component to the north-west, when imaged with the VLA in D-configuration. The FIR spectrum shows evidence for heated dust, with a cold dust temperature of $T_{\rm D} = 29.2^{+2.7}_{-2.3}$~K and an emissivity $\beta = 2.0\pm0.3$. These are consistent with heating due to star formation, with  $\mu_{\rm FIR}~\times~\rm{SFR} = 10^{3.6}$~M$_{\odot}$ yr$^{-1}$ \citep{Stacey2018}.

\section{Observations}
\label{sec:observations} 

In this section, we present our high resolution radio continuum and CO (1--0) observations, and the new and archival optical and NIR data that we used for our analysis.

\begin{table*}
	\centering
	\caption{Summary of the \textsl{Hubble Space Telescope} optical and Keck Adaptive-Optics NIR observations of MG~J0751+2716 and JVAS~B1938+666.
}
			\begin{tabular}{lllllll}
			\hline
			Target & Date  & Telescope & Instrument & Aperture/Camera & Filter & $t_{\rm exp}$ (s) \\
			\hline
		   \multirow{5}{*}{MG~J0751+2716} &  1999 May 13 & \textsl{HST} 	& WFPC2 		& PC1 		 & F555W 		& 1600 \\ 
			& 2000 Oct 21		 		 & \textsl{HST}   	& WFPC2 		& PC1 		 & F814W 		& 5200 \\
			& 1997 Oct 05        		 & \textsl{HST}	& NICMOS 	& NIC2 		 & F160W 		& 2560 \\
			& 2011 Dec 30 & Keck II 	& Nirc2 & Narrow  	& \textsl{K$^\prime$} & 3960 \\
      		& 2012 Dec 23 and 24   		 & Keck II 				& Nirc2 		& Narrow 	 & \textsl{K$^\prime$} 	& 7200 \\
			\hline
		 	\multirow{4}{*}{JVAS~B1938+666} &  1999 Apr 24   & \textsl{HST} 	 & WFPC2 	 & PC1	  	 & F555W 		&  2400\\
		 	& 1999 Apr 24  				 & \textsl{HST} 	 & WFPC2 	 & PC1	  	 & F814W 		&  3000\\
		 	& 1999 Apr 24  				 &\textsl{HST}		 & NICMOS	 & NIC2  	 & F160W 		& 5568\\\ 
			& 2010 Jun 29  &  Keck II & Nirc2   & Narrow & \textsl{K$^\prime$} & 15840\\
		\hline
	\end{tabular}
 \label{Tab:optical-obs}
\end{table*}

\begin{figure*}
\centering
    \includegraphics[width = 1.0\textwidth]{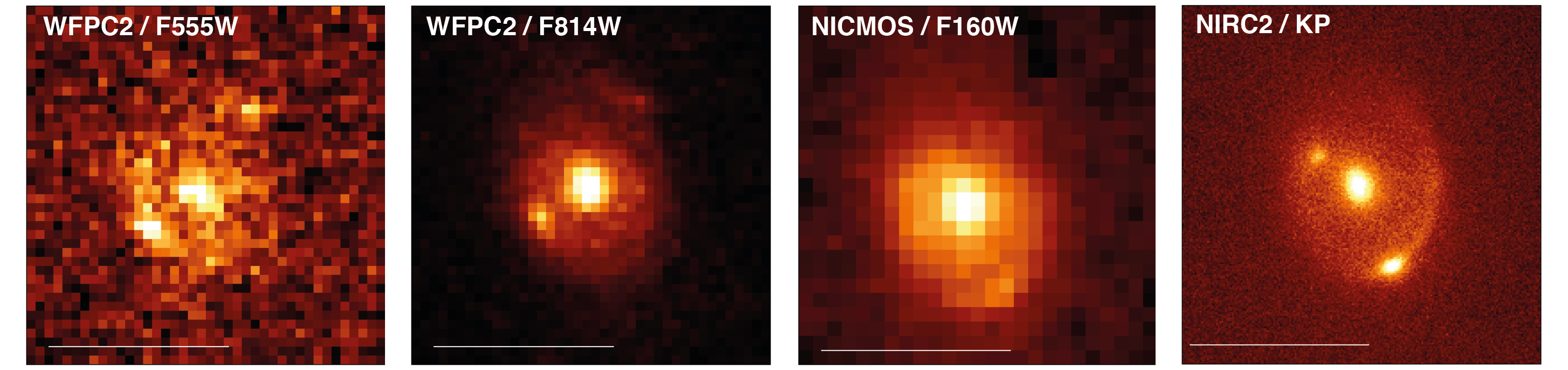}
    \caption{From left to right: optical and NIR imaging of MG~J0751+2716 taken with {\it HST}--WFPC2/F555W,{ \it HST}--WFPC2/F814W, {\it HST}--NICMOS/F160W and Keck-AO at 2.12~$\mu$m.  The white scale bar in each image represents 1~arcsec.
 }\label{Fig:0751-hst-keck}
\end{figure*}

\begin{figure*}
\centering
    \includegraphics[width = 1.0\textwidth]{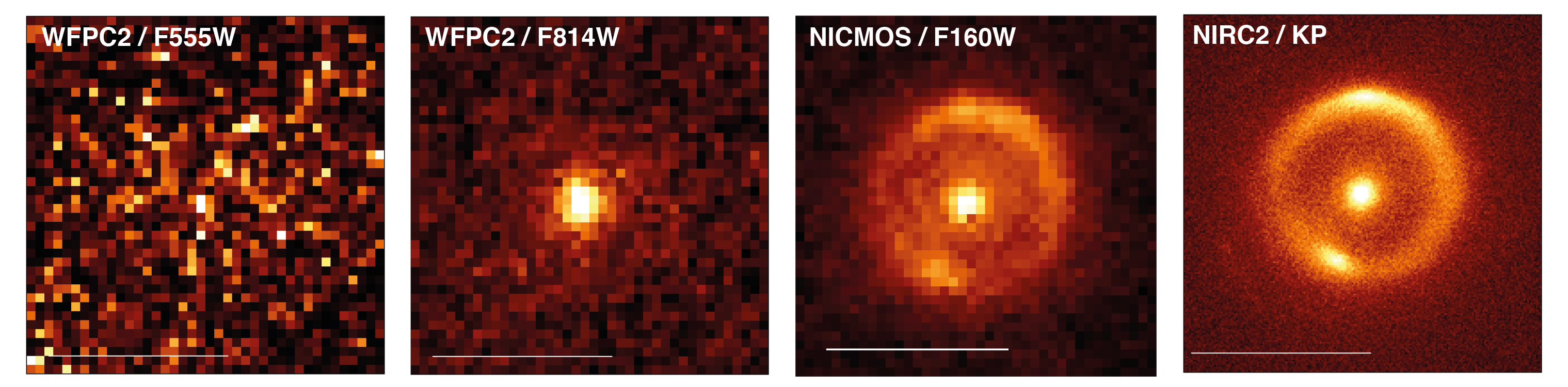}
\caption{From left to right: optical and NIR imaging of JVAS~B1938+666 taken with {\it HST}--WFPC2/F555W,{ \it HST}--WFPC2/F814W, {\it HST}--NICMOS/F160W and Keck-AO at 2.12~$\mu$m.  The white scale bar in each image represents 1~arcsec.
}\label{Fig:1938-hst-keck}
\end{figure*}

\subsection{W. M. Keck Telescope adaptive optics}
\label{sec:observations-keck}

MG~J0751+2716  and JVAS~B1938+666 were observed at 2.12~$\mu$m ($K^\prime$ band) with the adaptive-optics system on the W. M. Keck-II Telescope as part of the Strong-lensing at High Angular Resolution Programme (SHARP;  \citealt{McKean2007,Lagattuta2010,Lagattuta2012,Vegetti2012,Hsueh2016,Hsueh2017,Chen2016,Spingola2018,Chen2019,Hsueh2020}). The targets were observed using the NIRC2 instrument (narrow camera), which provides a field-of-view of 10~arcsec\,$\times$\,10~arcsec, and a pixel-scale of 9.942 mas~pixel$^{-1}$. The FWHM of the central part of the adaptive optics corrected point spread function (PSF) was about 65~mas. Further details of these observations are given in Table~\ref{Tab:optical-obs}. The data reduction was performed by following the methodology described by \citet{Auger2011}. The final 2.12~$\mu$m images for MG~J0751+2716 and JVAS~B1938+666 are presented in Figs.~\ref{Fig:0751-hst-keck} and \ref{Fig:1938-hst-keck}, respectively.

\subsection{\textit{Hubble Space Telescope}}
\label{sec:observations-hst}

High resolution optical/NIR observations of MG~J0751+2716 and JVAS B1938+666 that were taken with the {\it HST} were retrieved from the archive and re-processed. The datasets were mainly obtained as part of the CfA-Arizona Space Telescope Lens Survey (CASTLES; \url{www.cfa.harvard.edu/castles/}) and the Cosmic Lens All-Sky Survey (CLASS; \citealt{Browne2003, Myers2003}) programmes. For both targets, data were taken using the Wide-Field Planetary Camera 2 (WFPC2) through the F555W and F814W filters (GO-7495; PI: Falco, GO-8268; PI: Impey) and with the Near-Infrared Camera and Multi-Object Spectrograph (NICMOS) using the NIC2 camera through the F160W filter (GO-7255; PI: Jackson, GO-7495; PI: Falco). These data have been published most recently by \citet{Alloin2007} and \citet{Lagattuta2012}, and further details can be found there; an observational summary can also be found in Table~\ref{Tab:optical-obs}. The data were re-processed using {\sc astrodrizzle} within the {\sc iraf} package by applying standard procedures (see \citealt{Auger2009}) and using a final drizzled pixel scale of 50~mas~pixel$^{-1}$. The final reduced images from the {\it HST} are also presented in Figs.~\ref{Fig:0751-hst-keck} and \ref{Fig:1938-hst-keck} for MG~J0751+2716 and JVAS~B1938+666, respectively.

\subsection{Karl G. Jansky Very Large Array}
\label{sec:observations-jvla}

\begin{table}
	\centering
	\caption{Summary of the Ka-band observations of MG J0751+2716 and JVAS~B1938+666 with the VLA. 
    }
	\begin{tabular}{lllll}
		\hline
		Target & ID & Date & ConFig.~& $t_{\rm exp}$ (h) \\
		\hline
	 \multirow{10}{*}{MG J0751+2716} & 12A-319 &  2012 Jul 23 & B & 1.5 \\
		& 12A-319	&  2012 Aug 18 & B & 1.5\\
		& 12A-319	&  2012 Aug 29 & B & 1.5\\
		& 12A-319	&  2012 Aug 31 & B & 1.5 \\
		& 12A-319	&  2012 Sep 01 & B & 1.5  \\
		& 12A-319	&  2012 Sep 02 & B & 1.5 \\
		& 12A-319	&  2012 Sep 03 & B &  3 \\
		& 14B-301  & 2014 Oct 10  & C & 3 \\
		& 14B-301  & 2014 Oct 14  & C & 3 \\
		& 14B-301  & 2014 Oct 15  & C & 3 \\
		\hline
		\multirow{7}{*}{JVAS~B1938+666} & 11A-283 & 2011 Aug 05 & A & 2\\
		& 11A-283 & 2011 Aug 08 & A & 1.5\\
		& 11A-283 & 2011 Aug 05 & A &1.5\\
		& 12A-319 & 2012 Jun 05	 & B & 3\\
		& 12A-319 & 2012 Jun 08	 & B & 3\\
		& 12A-319 & 2012 Jun 09	 & B & 3\\
		& 15B-329 & 2015 Nov 06	 & D & 3\\
		\hline
	\end{tabular}\label{Tab:JVLA-obs}
\end{table}

\subsubsection{Observations}
We carried out high angular resolution interferometric observations that targeted the CO (1--0) emission line with the Ka-band receiver on the VLA, which used multiple configurations and frequency set-ups. MG J0751+2716 was observed using both the C- and B-configurations for 9 and 12~h, respectively, and JVAS~B1938+666 was observed using both the B- and A-configurations for 9 and 5~h, respectively (Project ID: 11A-283, 12A-319, 14B-301; PI: McKean). The data were typically taken in short 1.5 to 3~h observing blocks to aid the scheduling during good weather (see Table~\ref{Tab:JVLA-obs}).

The C- and B-configuration data were taken using 16 spectral windows with a bandwidth of 128 MHz each that were divided into 64 spectral channels; the total bandwidth for the observations was 2.048 GHz in a dual-polarization mode. The central observing frequency for the observations of MG~J0751+2716 was $\nu_{\rm obs} = 27.4455$~GHz, while for JVAS~B1938+666 it was $\nu_{\rm obs} = 37.6826$~GHz. The A-configuration data for JVAS~B1938+666 were taken using 2 spectral windows with 128 MHz bandwidth and 64 channels each using dual polarization. However, the spectral windows overlapped to reduce the effect of the sharp bandpass edges on any possible broad component of the emission line. A visibility averaging time of 3~s was used for all datasets. In addition, we re-analyzed archival VLA observations in D-configuration for JVAS~B1938+666 (Project ID: 15B-329, PI: Sharon). These observations were for 3 h in total, and were taken using 16 spectral windows with a bandwidth of 128 MHz each. Note that 8 spectral windows covering frequencies that included the line emission were divided into 128 spectral channels, whereas 64 channels were used for the 8 spectral windows that covered the continuum emission (see \citealt{Sharon2016}). 

The observing strategy for the observations were the same for all configurations; 3C48, 3C286 or 3C147 were used for the absolute flux-density calibration and nearby phase reference sources were used to determine the relative antenna gains (amplitude and phase) as a function of time and frequency, and also to check the antenna pointing every 1.5~h, when required. The scans on the target were $\sim 3.5$~min each, which were interleaved by $\sim 1.5$~min scans on the phase-reference calibrator. This is a longer cycle-time than is recommended for such high frequency observations with the VLA. However, given the strong continuum flux-density and large bandwidths that were used, we were able to use self-calibration of the targets to determine the phase variations on shorter time-scales (see below).

\subsubsection{Calibration method}

All of the datasets were reduced with the Common Astronomy Software Application package ({\sc casa}; \citealt{McMullin2007}) using scripts that applied standard calibration procedures. Here, we summarize the steps. We first perform the standard a-priori calibrations, which included corrections for the antenna positions, the tropospheric opacity and the antenna gain curves as a function of elevation. We then inspect the visibilities in order to flag potential bad data. For the observations of MG~J0751+2716, we found radio frequency interference (RFI) between 27.4 and 27.9 GHz that strongly affected the visibilities in the LL circular polarization. Therefore, we use only the visibilities in the RR polarization for the MG~J0751+2716 dataset. We generally had to flag 2 to 3 antennas out of 27 antennas per observation that were not operational, and the initial 21~s of each scan, when the antennas were recording data while still slewing on-source.

To solve the radio-interferometric Measurement Equation, we use the pre-determined models for the primary flux-density calibrators that have been made by \citet{Perley2013}, adjusted to the correct frequency, and assume a point-source model (flat spectrum) for the secondary phase calibrator. Using these models for the primary calibrator, we solve for the antenna-based delays (phase as a function of frequency) and determine the bandpass solutions (amplitude as a function of frequency). We then determine the amplitude and phase solutions as a function of time using both the primary and secondary calibrators. This process provides a new model for the secondary calibrator, which has the correct flux density and broad-band spectrum. This new model is then used to determine the antenna complex gains as a function of time and frequency. Additional flagging is then done, if required, and the process is repeated until the residual visibilities (calibrated -- model) show no major outliers anymore. Finally, we apply the calibration solutions to each calibrator, and then to the target source by interpolating the solutions from the secondary calibrator. 

\subsubsection{Imaging and self-calibration of the continuum emission}

\begin{figure*}
\centering
	\includegraphics[width = 0.48\textwidth]{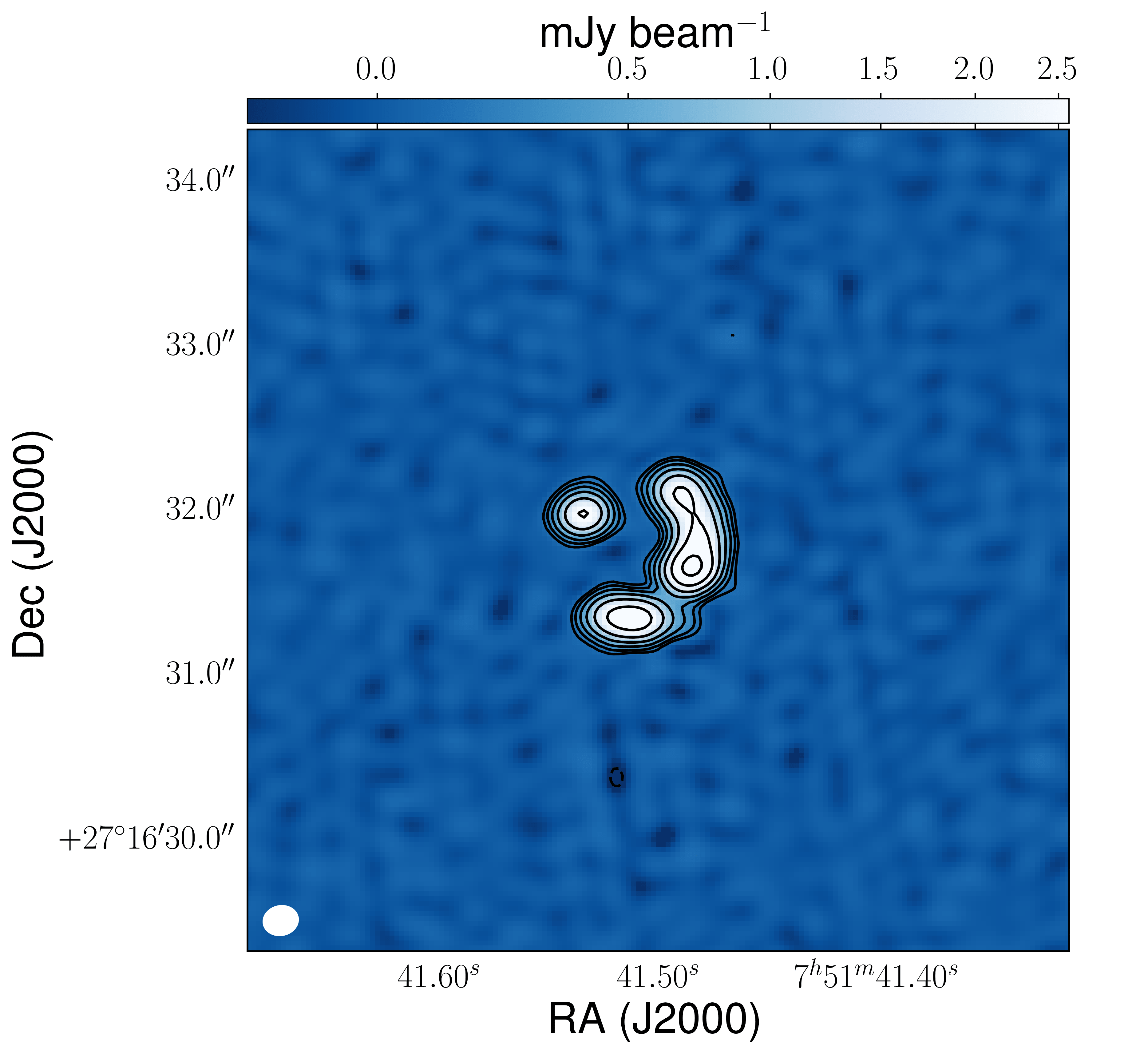}
	\includegraphics[width = 0.48\textwidth]{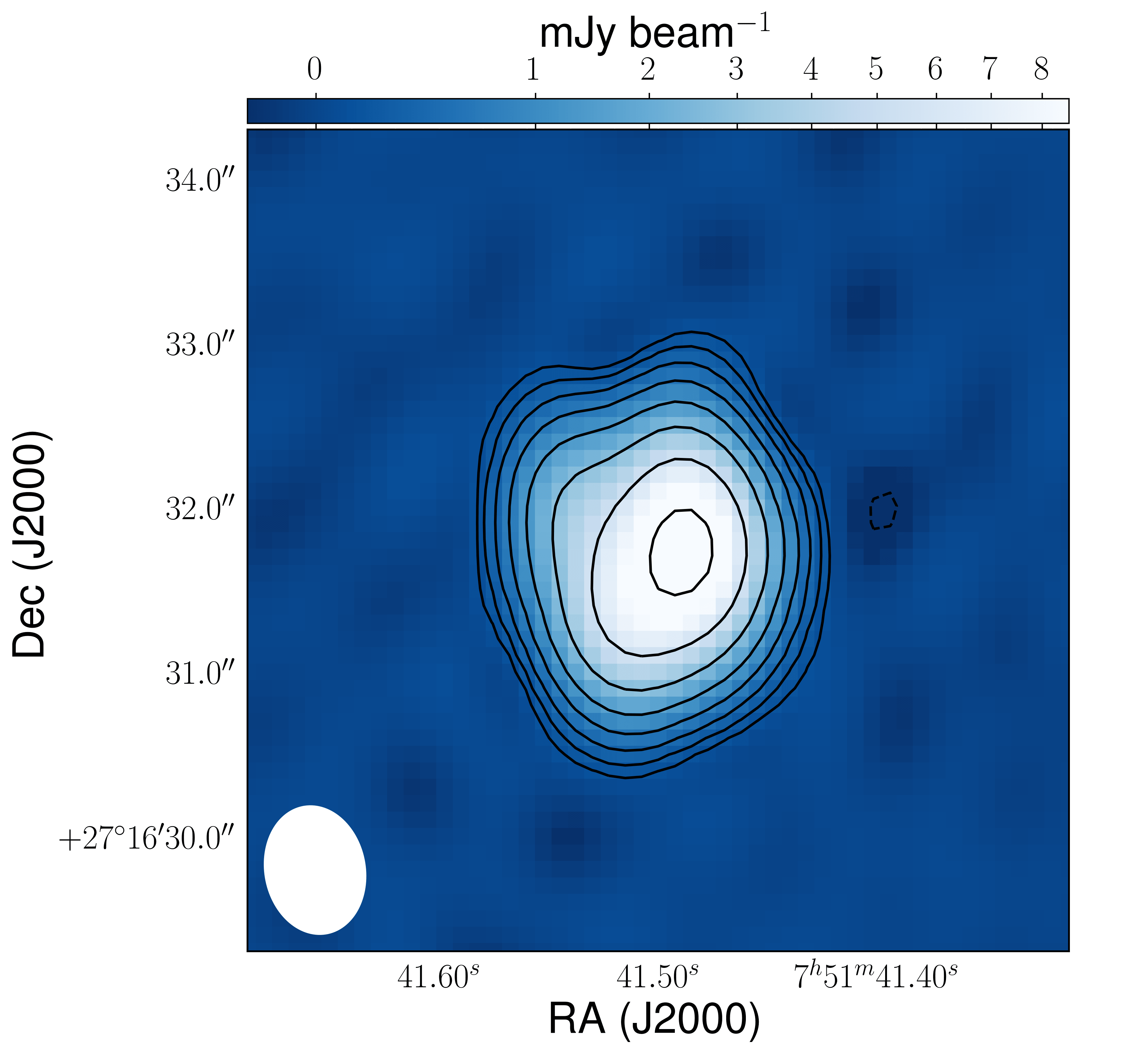}
    \caption{The 27.4 GHz VLA continuum image of MG J0751+2716, taken in B-configuration (left) and C-configuration (right). For the B-configuration image, the off-source rms noise level is $22~\mu$Jy~beam$^{-1}$ and the peak surface brightness is 6~mJy~beam$^{-1}$. The synthesized beam is shown in the bottom left corner and is $0.21 \times 0.18$~arcsec$^2$ at a position angle of $78.15$~deg east of north. For the C-configuration image, the off-source rms noise level is $19~\mu$Jy~beam$^{-1}$ and the peak surface brightness is 12~mJy~beam$^{-1}$. The synthesized beam is shown in the bottom left corner and is $0.78 \times 0.61$~arcsec$^2$ at a position angle of $-169.5$~deg east of north. In both images, the first contour is 3 times the off-source rms noise level and the contour levels increase by a factor of 2. }
\label{Fig:0751-jvla-continuum}
\end{figure*}

\begin{figure*}
\centering
	\includegraphics[width = 0.48\textwidth]{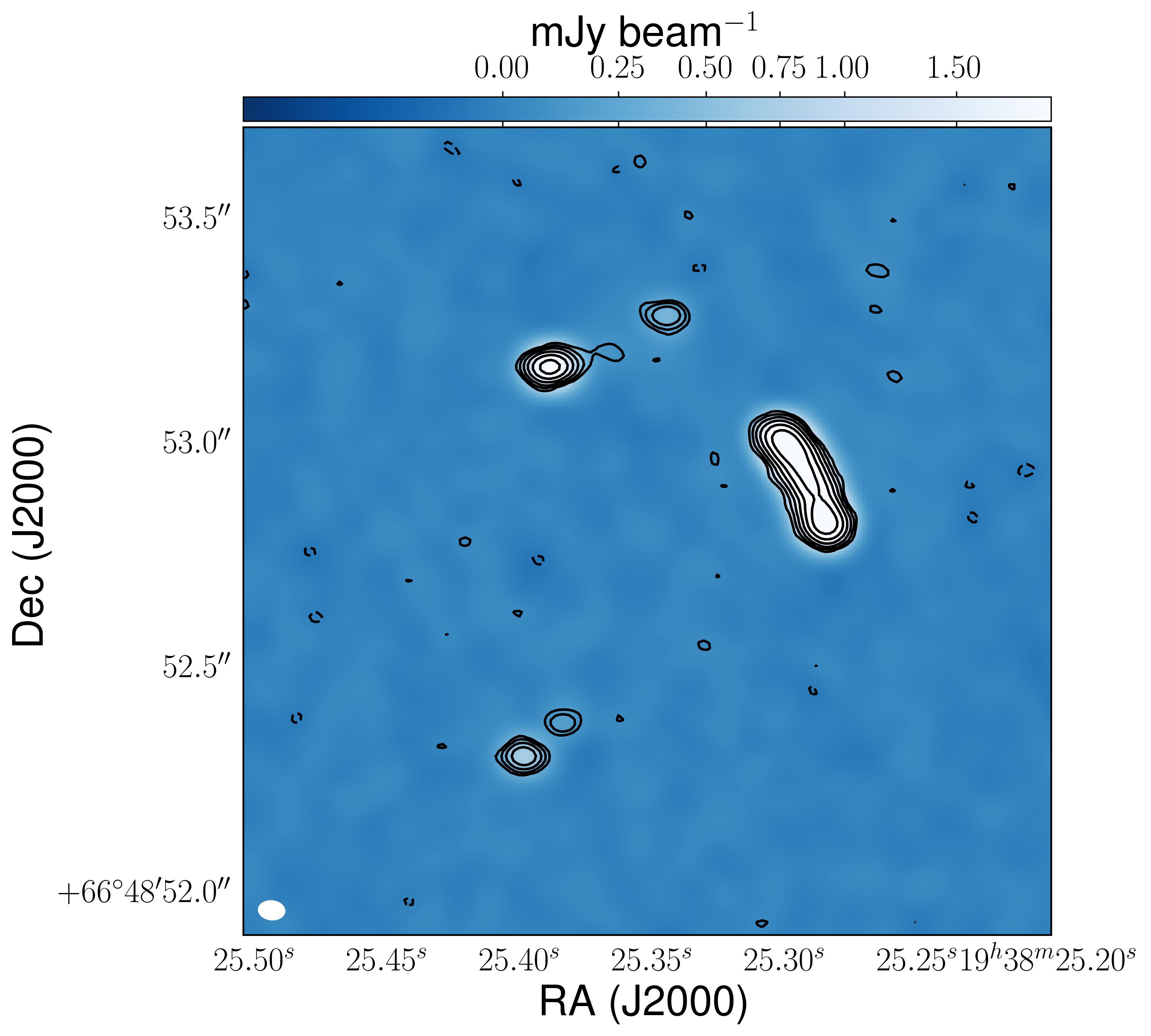}
    \includegraphics[width = 0.48\textwidth]{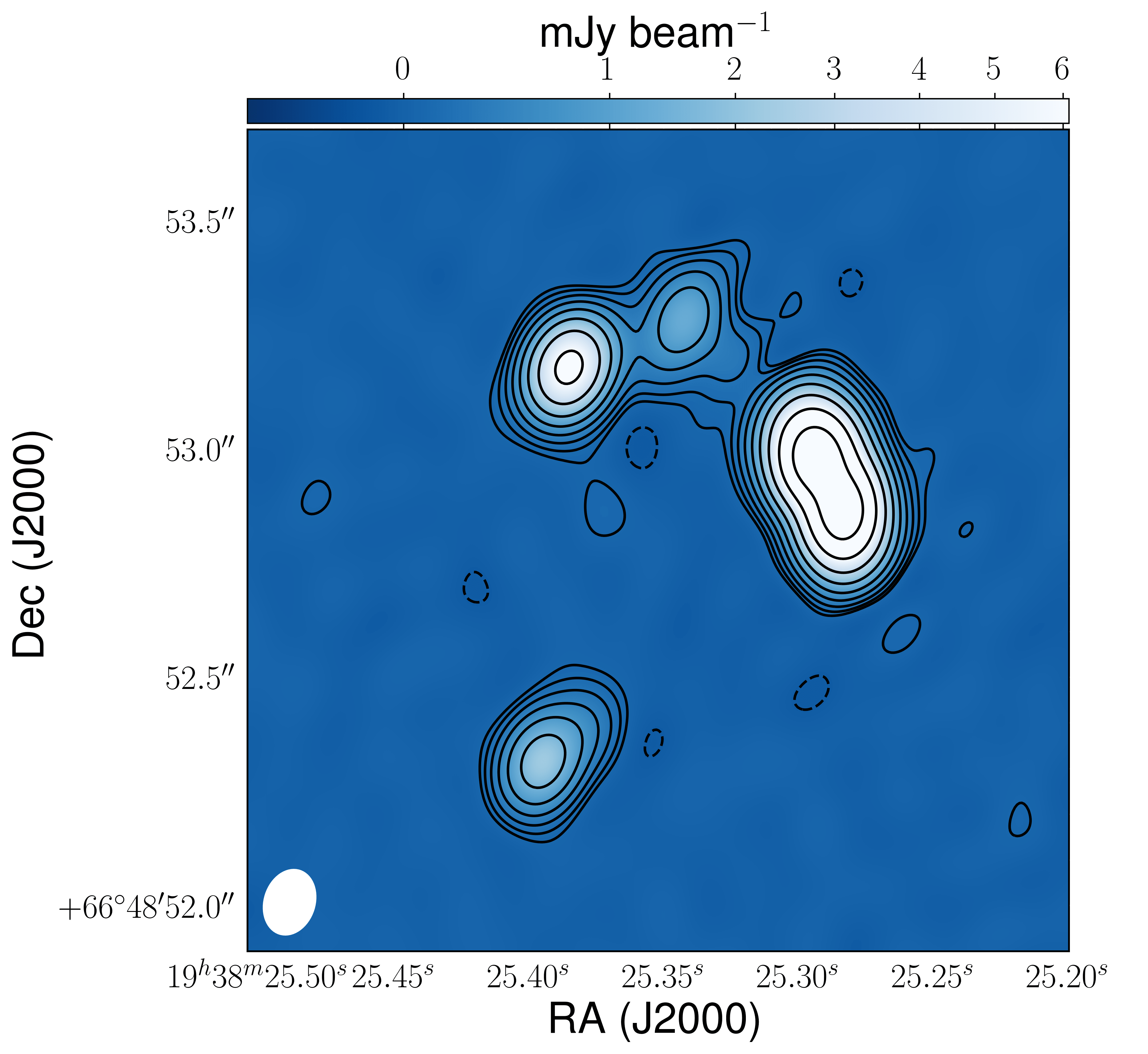}
    \caption{The 37.7 GHz VLA continuum image of JVAS~B1938+666, taken in A-configuration (left) and B-configuration (right). For the A-configuration image, the off-source rms noise level is $40~\mu$Jy~beam$^{-1}$ and the peak surface brightness is 9~mJy~beam$^{-1}$. The synthesized beam is shown in the bottom left corner and is $0.058 \times 0.041$~arcsec$^2$ at a position angle of $84.05$~deg east of north. For the B-configuration image, the off-source rms noise level is $17~\mu$Jy~beam$^{-1}$ and the peak surface brightness is 20~mJy~beam$^{-1}$. The synthesized beam is shown in the bottom left corner and is $0.15 \times 0.11$~arcsec$^2$ at a position angle of $-17.9$~deg east of north. In both images, the first contour is 3 times the off-source rms noise level and the contour levels increase by a factor of 2.}
\label{Fig:1938-jvla-continuum}
\end{figure*}

We perform phase-only self-calibration for the continuum emission separately for each dataset using the line-free spectral windows. We first make an image based on the calibration procedure described above and then use this model for the source to determine the phase solutions as a function of time. We start with a solution interval that is equal to the scan length, and then decrease this iteratively to a solution interval of 60~s. Note that during this process we apply the phase solutions to all spectral windows, including those with the spectral line. We then concatenate all of the self-calibrated datasets in order to perform the final imaging. We do not use the edge channels of the spectral windows, which are often noisier than the central channels. The images are obtained using a Briggs weighting scheme \citep{Briggs1995} with the robust parameter set to 0, which is a good compromise between short and long baseline weighting for the VLA. The phase-only self-calibrated images of MG~J0751+2716 and JVAS~B1938+666 are presented in Figs.~\ref{Fig:0751-jvla-continuum} and \ref{Fig:1938-jvla-continuum}, respectively.

\subsubsection{Continuum subtraction and spectral line imaging}

The first step in extracting the CO (1--0) emission consists of subtracting the continuum emission from the interferometric data. This is done by fitting a first order polynomial function in the Fourier space to the real and imaginary parts of the line-free spectral windows only. The model of the continuum is then subtracted from the visibilities and a continuum-subtracted dataset is generated. We then made a dirty cube image in order to estimate the rms noise per channel. Subsequently, we interactively {\sc clean} the cube using a threshold that is 3 times the noise of the line-free channels and a natural weighting scheme. We also made cubes by applying a {\it uv}-taper and also using Hanning smoothing of the {\it uv}-data (at the expense of reducing the velocity resolution by a factor of two) to enhance the extended emission from the molecular gas and improve the de-convolution of the channel data. In addition, we combined the visibility data from different VLA configurations for both targets. However, as the highest angular resolution observation in A-configuration for JVAS B1938+666 and in B-configuration for MG~J0751+2716 did not detect the CO (1--0) emission line (see below and Fig.~\ref{Fig:appendix}), the surface brightness sensitivity, and hence, the significance of our detections went down for the combined datasets. Therefore, for our analysis, we use only the B- and C-configuration datasets for JVAS B1938+666 and MG~J0751+2716, respectively.

\begin{figure}
\centering
	\includegraphics[width = 0.48\textwidth]{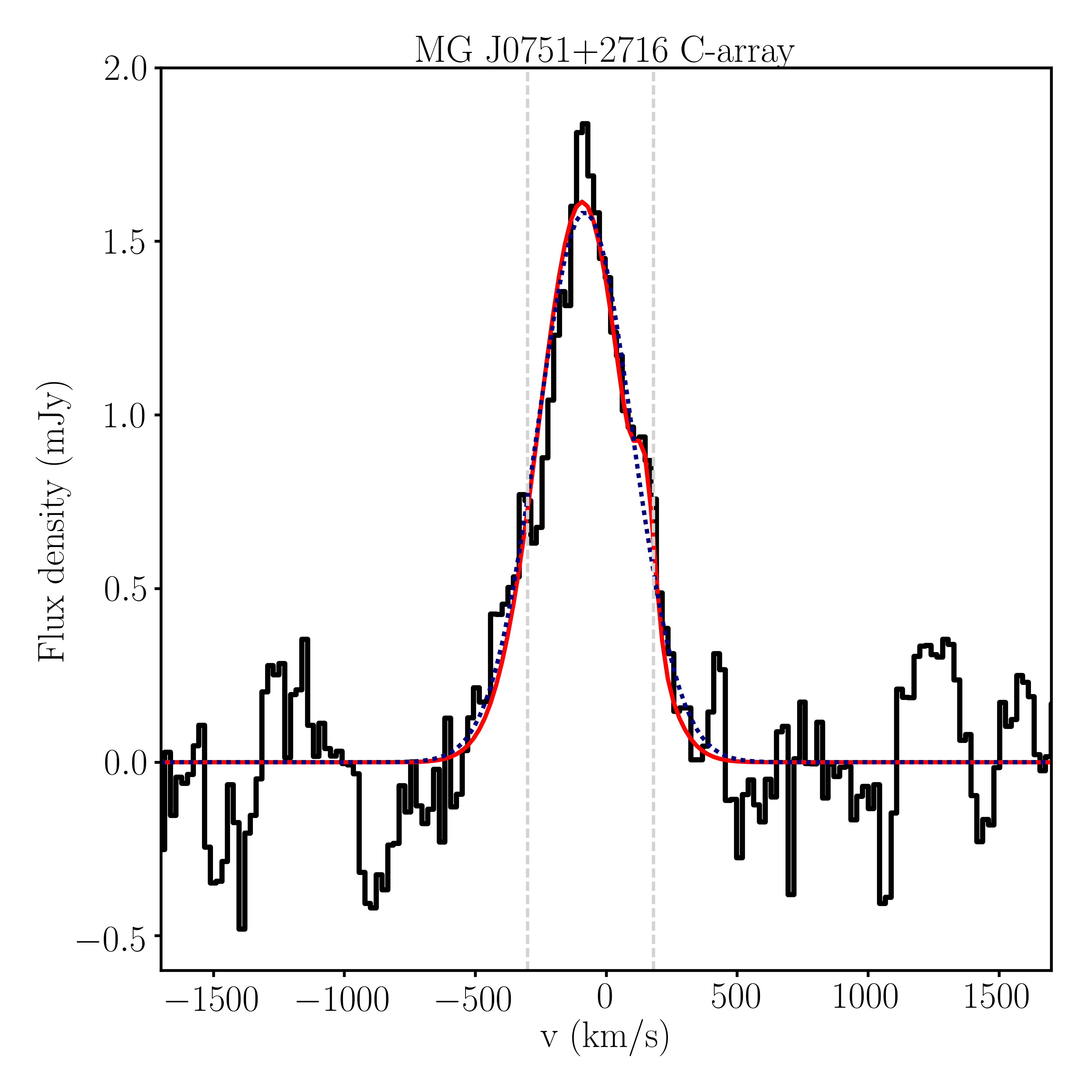}
    \caption{The integrated spectrum of the CO (1--0) molecular gas emission from MG J0751+2716 from the C-configuration data, relative to the systemic velocity for $z = 3.200$, using the radio definition of the velocity ($\nu_{\rm sys} = 27.44552$~GHz). The dotted blue line is the best single Gaussian fit, while the red line is the best double Gaussian fit. The dashed vertical lines indicate the velocity range used for the moment maps ($-300$ to $+180$~km\,s$^{-1}$). The velocity width of each channel is 21.8~km\,s$^{-1}$ and the spectrum has been smoothed with a boxcar of 109~km\,s$^{-1}$ (5 channels) width for clarity.} \label{Fig:0751-spectrum}
\end{figure}

\begin{figure}
\centering
	\includegraphics[width = 0.48\textwidth]{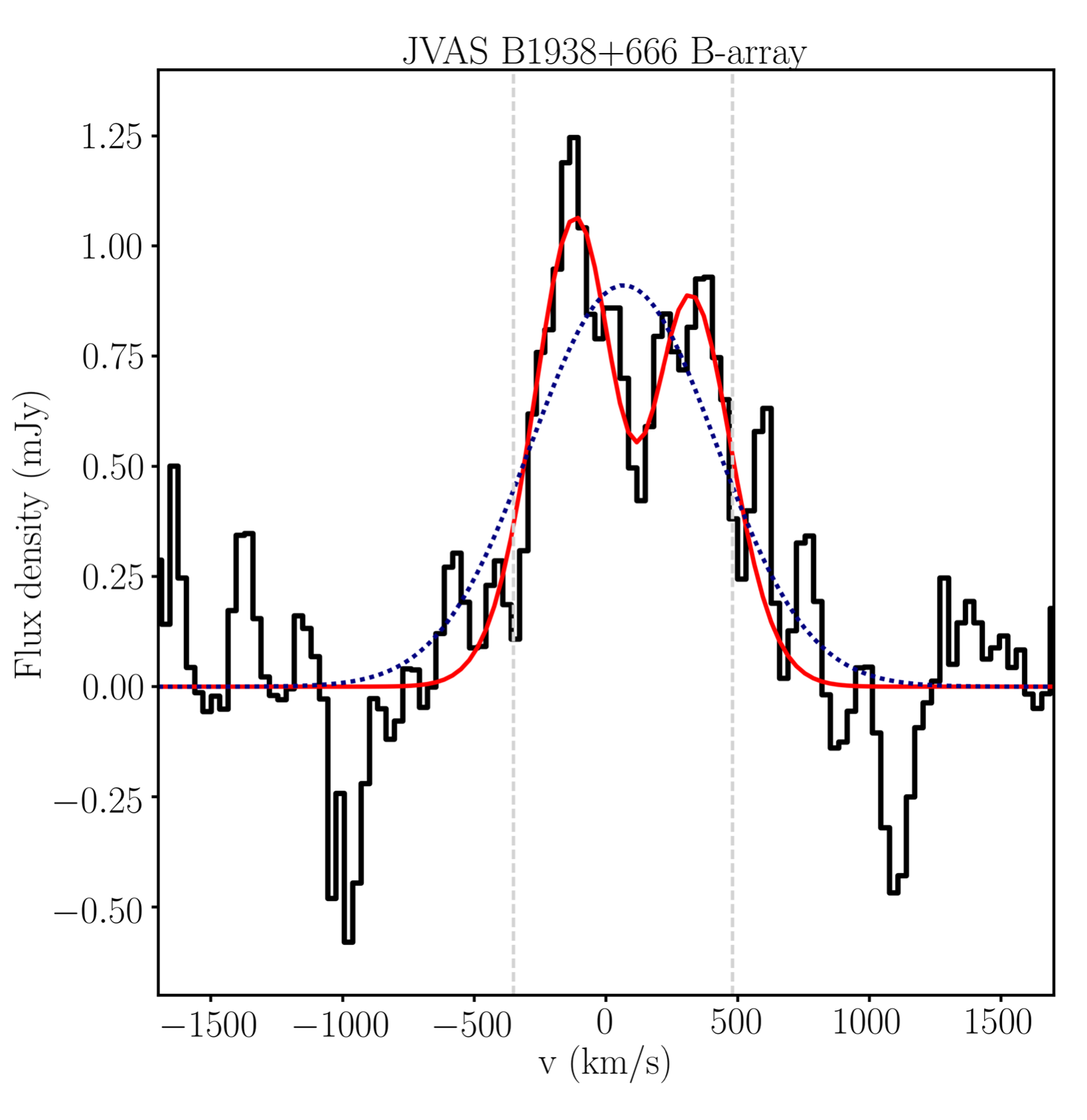}
    \caption{The integrated spectrum of the CO (1--0) molecular gas emission from JVAS~B1938+666 from the B-configuration data, relative to the systemic velocity for $z = 2.059$, using the radio definition of the velocity ($\nu_{\rm sys} = 37.68264$~GHz). The dotted blue line is the best single Gaussian fit, while the red line is the best double Gaussian fit.  The dashed vertical lines indicate the velocity range used for the moment maps ($-350$ to $+480$~km\,s$^{-1}$). The velocity width of each channel is 31.8~km\,s$^{-1}$ and the spectrum has been smoothed with a boxcar of 95.4~km\,s$^{-1}$ (3 channels) width for clarity.}
\label{Fig:1938-spectrum}
\end{figure}

\begin{figure*}
\centering
	\includegraphics[width = 0.855\textwidth]{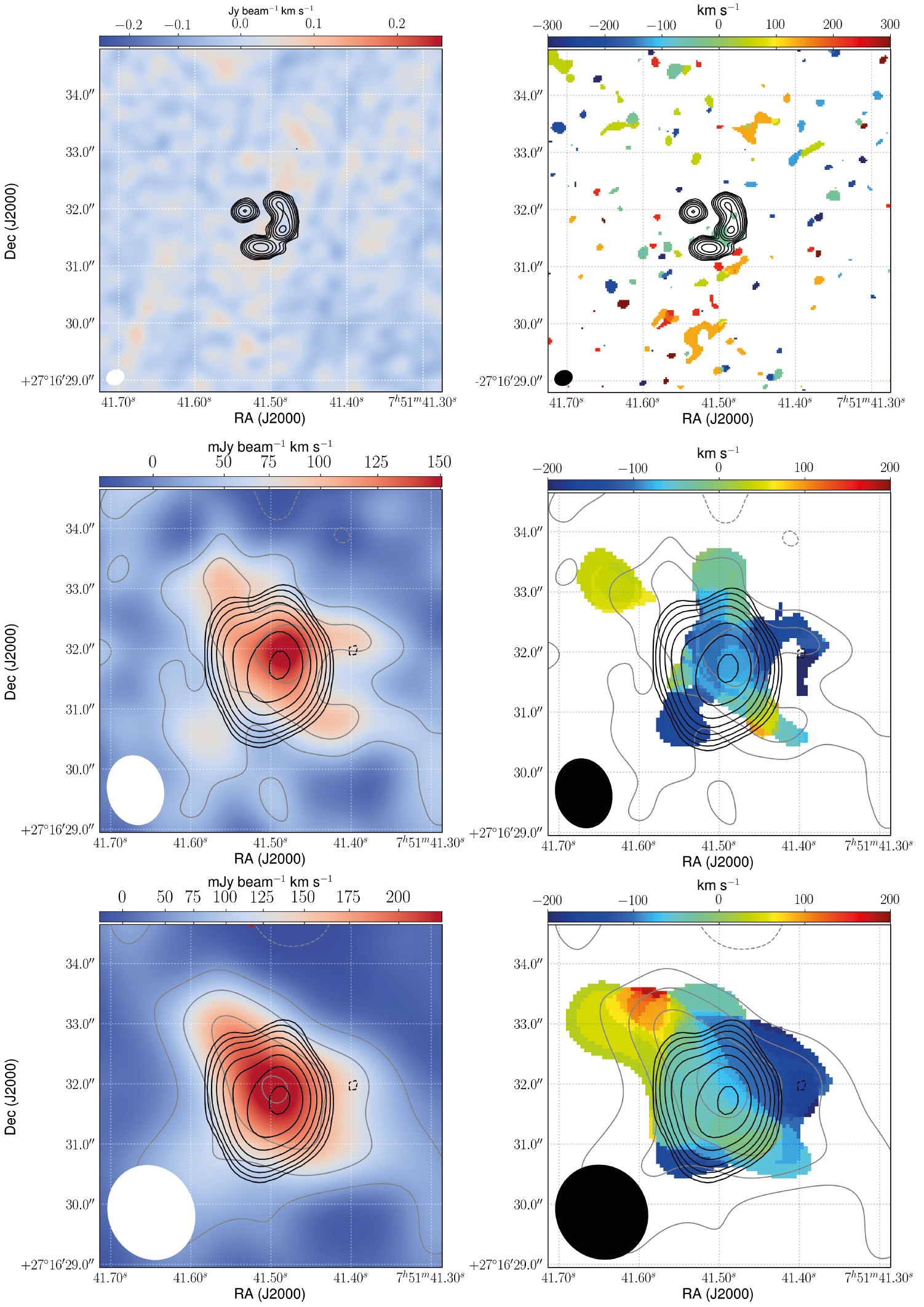}
    \caption{ The moment zero (left, obtained in the velocity range $-300$ to $+180$~km\,s$^{-1}$) and moment one (right) maps of the CO (1--0) emission from MG~J0751+2716 for B-configuration (first row), C-configuration using natural weighting (second row) and with a 1 arcsec FWHM Gaussian taper of the {\it uv}-data (third row). The contours represent the B-configuration and C-configuration continuum emission for reference, and are in successive integer powers of two times the off-source rms noise (30 $\mu$Jy beam$^{-1}$). Note that there is no detection of the CO (1--0) emission in the observations in B-configuration. The moment maps have been made using cubes with channel widths of 43.7~km\,s$^{-1}$. The synthesized beam is shown in the bottom left corner of each panel and is $0.212 \times 0.178$~arcsec$^2$ at a position angle of $-$78~deg (first row), $1.16 \times 0.93$~arcsec$^2$ at a position angle of 12.5~deg (second row), and $1.60 \times 1.42$~arcsec$^2$ at a position angle of 25.5~deg (third row). In each image we also overplot the moment zero contours (white in the left column, grey in the right column), which increase by factors of two. The first contour level is the off-source rms, which is  23 mJy~km\,s$^{-1}$~beam$^{-1}$ for the naturally weighted map and 30 mJy~km\,s$^{-1}$~beam$^{-1}$ for the tapered map.}
    \label{Fig:0751-moments}
\end{figure*}

\begin{figure*}
\centering
\includegraphics[width = 0.95\textwidth]{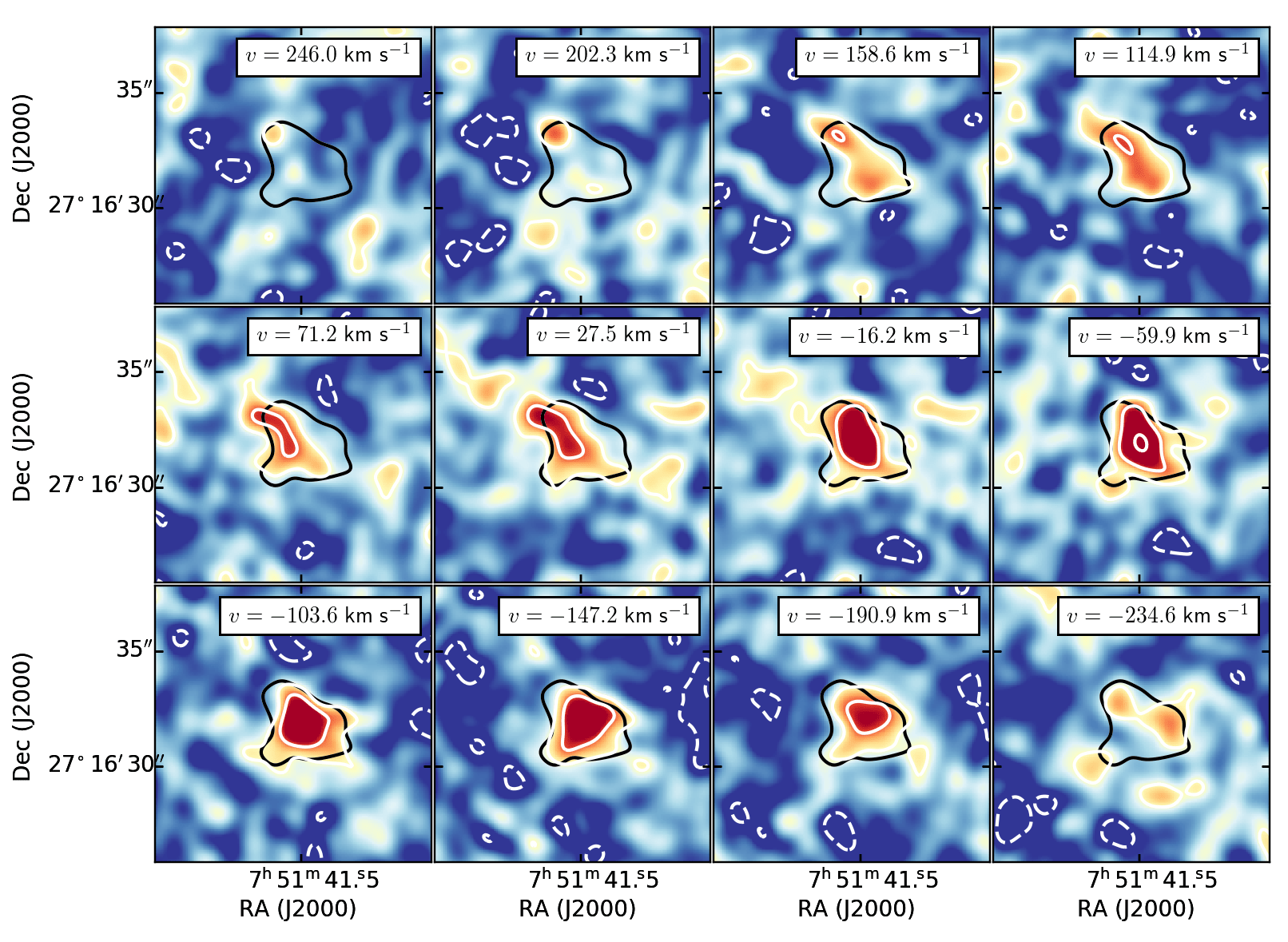}
    \caption{Channel maps (43.7 km\,s$^{-1}$ width) for the CO (1--0) emission line from $-235$ to 246~km\,s$^{-1}$ in MG~J0751+2716, relative to the systemic velocity for $z = 3.200$, using the radio definition of the velocity ($\nu_{\rm sys} = 27.44552$~GHz). The velocity of each channel map is indicated at the top right of each panel.  The black contours represent the moment-zero emission imaged using natural weights at $3 \times \sigma_{\rm rms}$, which is 30~mJy~km\,s$^{-1}$ beam$^{-1}$. The white contours increase by factors of two, where the first contour corresponds to the off-source rms noise per channel, which is on average 0.1~mJy~beam$^{-1}$. The data have been Hanning smoothed and the cube is tapered to give a lower resolution image with a synthesized beam of $1.16 \times 1.42$~arcsec$^2$ at a position angle of 25.4~deg east of north. The linear intensity scale ranges from $-0.1$ to 0.7~mJy~beam$^{-1}$~km\,s$^{-1}$. The maps show that the CO (1--0) emission is extended with a velocity dependent structure.}
    \label{Fig:0751-channel-map}
\end{figure*}

\begin{figure*}
\centering
\includegraphics[width = 0.95\textwidth]{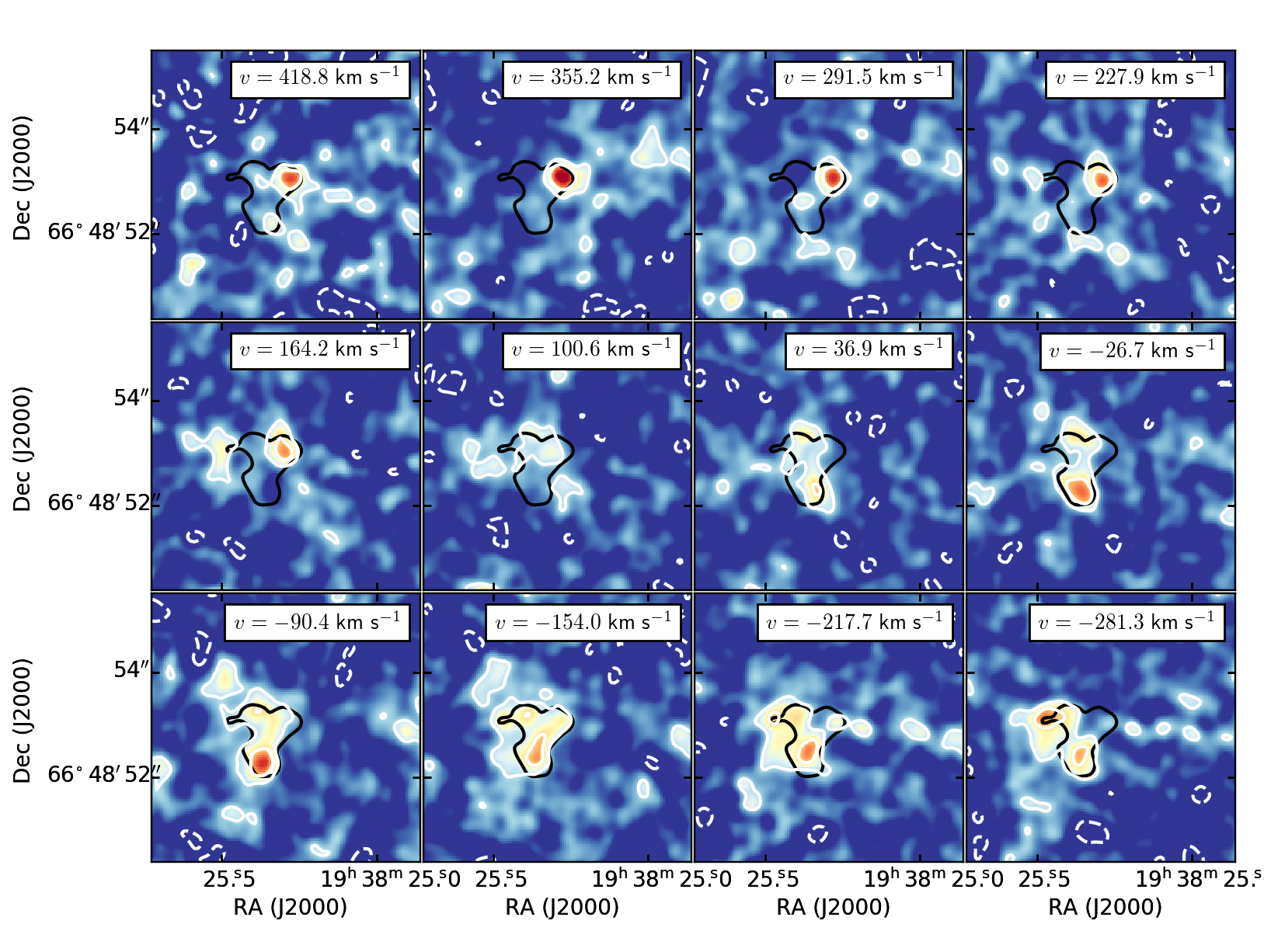}
    \caption{Channel maps (63.6 km\,s$^{-1}$ width) for the CO (1--0) emission line from $-281.3$ to 418.8~km\,s$^{-1}$ in JVAS~B1938+666, relative to the systemic velocity for $z = 2.059$, using the radio definition of the velocity ($\nu_{\rm sys} = 37.68264$~GHz). The velocity of each channel map is indicated at the top right of each panel. 
     The black contours represent the moment-zero emission at $3 \times \sigma_{\rm rms}$, which is 43~mJy~km\,s$^{-1}$ beam$^{-1}$. The white contours increase by factors of two, where the first contour corresponds to the off-source rms noise per channel, which is on average 75~$\mu$Jy~beam$^{-1}$.
    The data have been Hanning smoothed and the cube is tapered to give a lower resolution image with a synthesized beam of $0.42  \times  0.41$~arcsec$^2$ at a position angle of 4.9~deg east of north. The linear intensity scale ranges from $-0.01$ to 0.5~mJy~km\,s$^{-1}$~beam$^{-1}$. The maps show that the CO (1--0) emission is extended with a velocity dependent structure.}
    \label{Fig:1938-channel-map}
\end{figure*}

The spectra of the CO (1--0) emission, integrated over the extent of both lens systems, are presented in Figs.~\ref{Fig:0751-spectrum} and \ref{Fig:1938-spectrum} for MG J0751+2716 and JVAS~B1938+666, respectively. Finally, we construct moment maps for the integrated line intensity (moment zero) and for the line-intensity-weighted velocity (moment one), which are shown in Figs.~\ref{Fig:0751-moments} and \ref{Fig:1938-moments}, and will be described in detail in Section~\ref{sec:results-imageplane}. The channel maps are shown in Figs.~\ref{Fig:0751-channel-map} and \ref{Fig:1938-channel-map}. For the moment maps, we use those channels within the FWHM of the line profiles (see Figs.~\ref{Fig:0751-spectrum} and \ref{Fig:1938-spectrum}) and for the moment one map, we apply a signal-to-noise ratio cut of $4\sigma$.  We do not apply any aperture to the moment zero maps. Unfortunately, the signal-to-noise ratio was too low to make a robust velocity dispersion map (moment two). 

\begin{figure*}
\centering
\includegraphics[width = 0.95\textwidth]{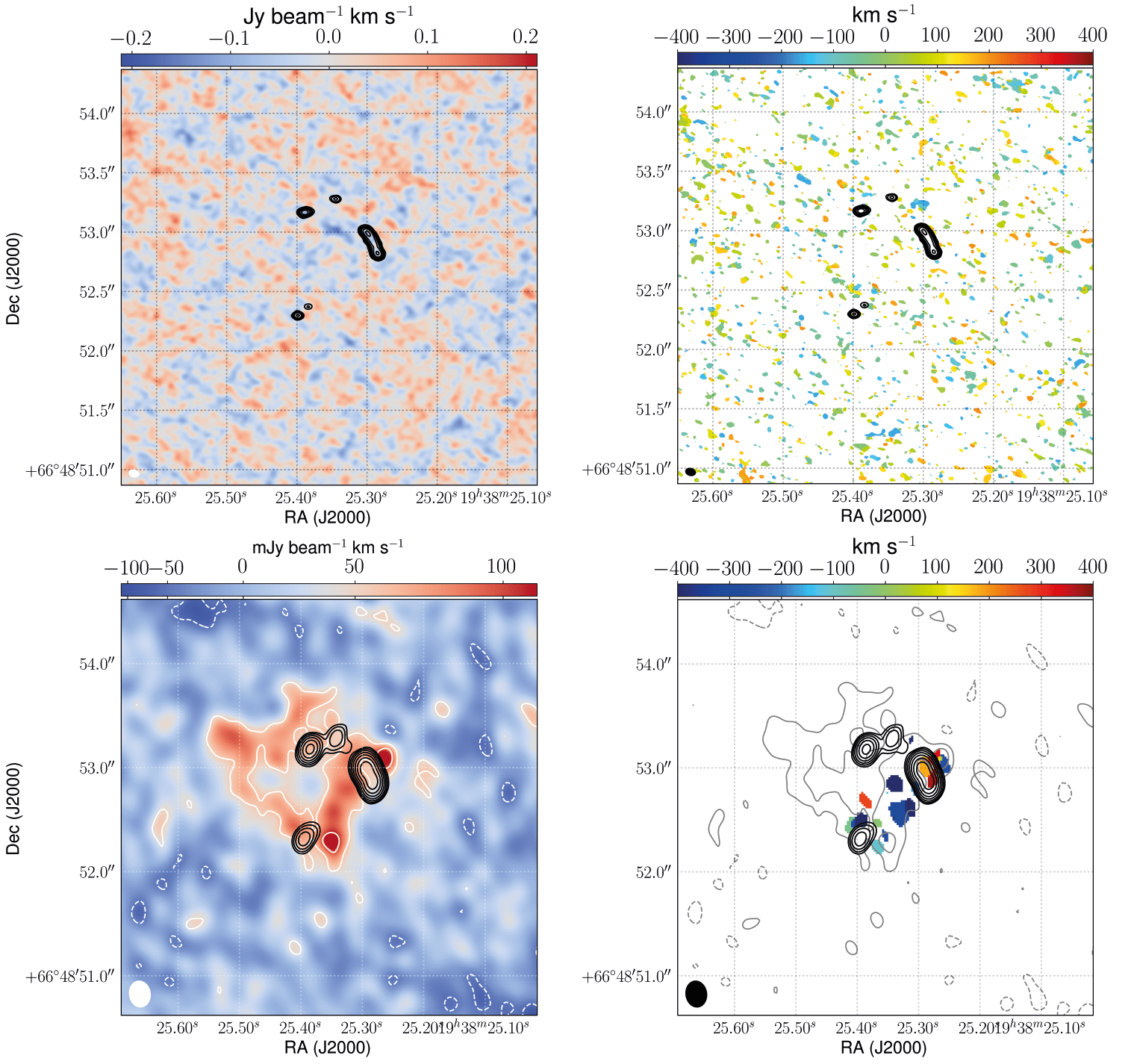}
    \caption{The moment zero (left,  obtained in the velocity range $-350$ to $+480$~km\,s$^{-1}$) and moment one (right) maps of the CO (1--0) emission from JVAS~B1938+666 for A-configuration using natural weighting (first row), B-configuration using natural weighting (second row) and with a 0.3 arcsec FWHM Gaussian taper of the {\it uv}-data (third row), and D-configuration using natural weighting (fourth row). The moment maps have been made using cubes with channel widths of 31.8~km\,s$^{-1}$. The black contours represent the A-configuration (first row) and B-configuration (second to fourth row) continuum emission for reference. Contours increase by powers of two; the first contour is the off-source rms noise level (40 and 60~$\mu$Jy beam$^{-1}$ for A-configuration and B-configuration, respectively). Note that there is no detection of the CO (1--0) emission in the observations in A-configuration. The synthesized beam is shown in the bottom left corner of each panel and is $0.085 \times 0.061$~arcsec$^2$ at a position angle of 77~deg (first row), $0.25 \times 0.20$~arcsec$^2$ at a position angle of 8.5~deg (second row), $0.42 \times 0.41$~arcsec$^2$ at a position angle of 4.9~deg (third row) and $2.6 \times 1.9$~arcsec$^2$ at a position angle of $-$46.1~deg (fourth row).  In each image we also overplot the moment zero contours (white in the left column, grey in the right column), which increase by factors of two. The first contour level corresponds to the off-source rms, which is 30 mJy~km\,s$^{-1}$~beam$^{-1}$ for the naturally weighted image (second row), 43 mJy~km\,s$^{-1}$~beam$^{-1}$ for the tapered image (third row) and 47 mJy~km\,s$^{-1}$~beam$^{-1}$ for the D-array image (fourth row).}
    \label{Fig:1938-moments}
\end{figure*}
\begin{figure*}
\centering
\includegraphics[width = 0.95\textwidth]{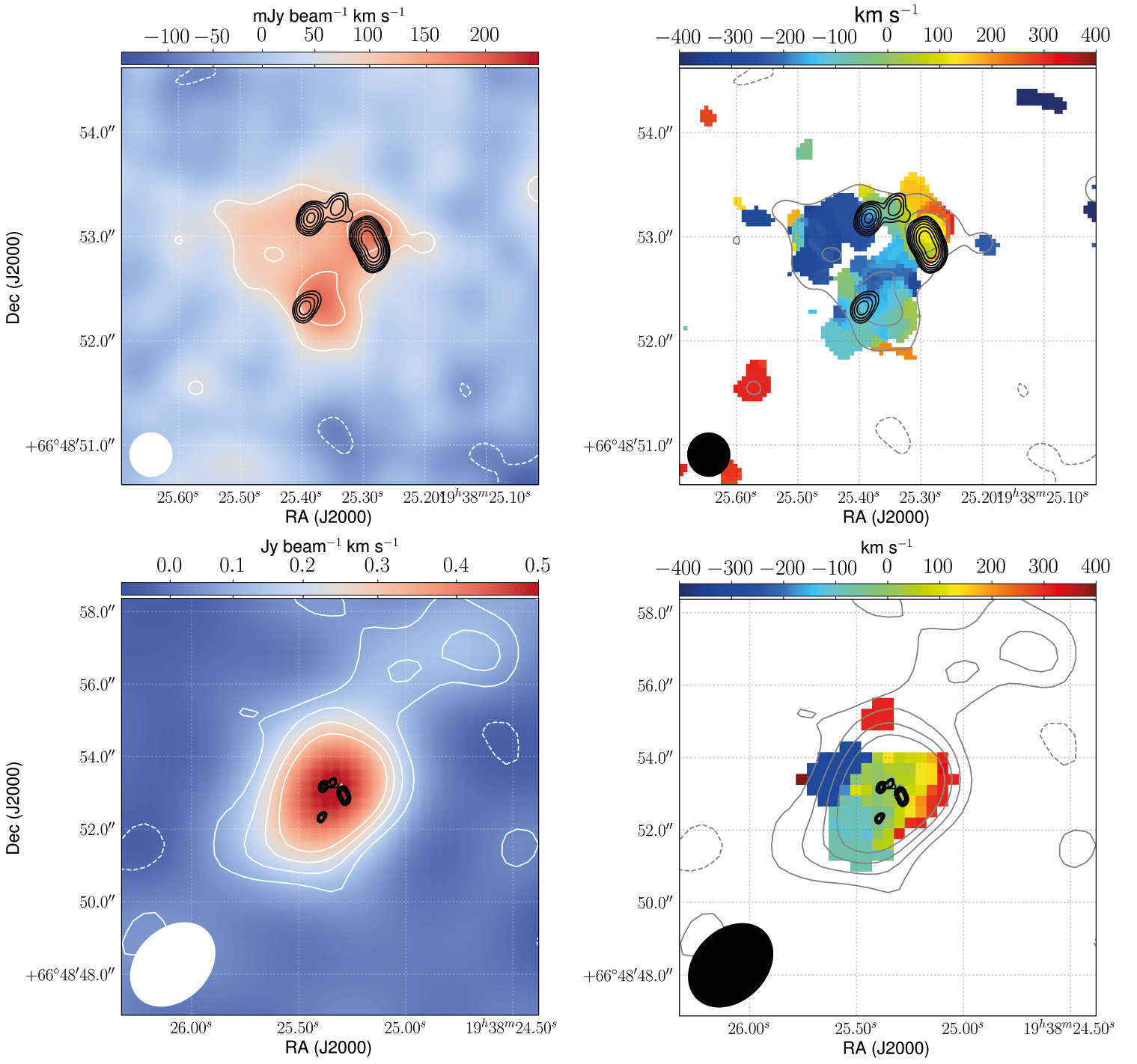}
\contcaption{}
\end{figure*}


\section{Lens Plane properties}
\label{sec:results-imageplane}

We now review the multi-wavelength properties of MG~J0751+2716 and JVAS~B1938+666 in the lens plane, which are summarized in Fig.~\ref{Fig:RGB}, and we also compare with previous results taken at a lower angular resolution.

\subsection{MG~J0751+2716}  

At the observed-frame NIR wavelengths (2.12 $\mu$m) the lensed source in MG~J0751+2716 consists of two compact red components with a faint extended arc that is due to the stellar emission from the AGN host galaxy (see Figs.~\ref{Fig:0751-hst-keck} and \ref{Fig:RGB}). We note that the improved resolution and surface brightness sensitivity provided by the adaptive optics imaging from the SHARP survey clearly resolves the AGN host galaxy emission in both lensed images from that of the foreground lensing galaxy (see Fig.~\ref{Fig:0751-hst-keck} fourth column). As already discussed by \citet{Alloin2007}, at optical wavelengths there are also two blue components that are within the Einstein radius of the system. The nature of these two blue components is unclear, as there is currently no redshift information available for them, and they have similar colours to the blue group members of the main lens \citep{Spingola2018}. However, as we will show in the next section, their surface brightness distribution and image geometry are consistent with a multiply imaged blue star-forming companion at the same redshift as the AGN host galaxy. 

The radio continuum emission of the background source is slightly resolved in our VLA C-configuration imaging, but is well-resolved in the B-configuration imaging (see Fig.~\ref{Fig:0751-jvla-continuum}); these data show two extended arcs and a slightly resolved component at 27.4 GHz that is consistent with previous imaging at higher angular resolution and at a lower-frequency \citep{Lehar1997, Spingola2018}. We find that the emission detected on VLBI-scales by \citet{Spingola2018} has a faint compact component that is coincident with the centre of the AGN host galaxy, but the dominant emission is offset and is presumably due to the non-thermal jet emission that extends beyond the host galaxy's stellar halo. We believe the radio emission coincident with the AGN host galaxy to be the AGN core. However, higher angular resolution observations with spectral index information are necessary to confidently identify the radio core (expected to have a flat radio spectral index) in this system. The total continuum flux density is $S_{\rm 27.4~GHz} = 24\pm2$~mJy and $S_{\rm 27.4~GHz} = 25\pm3$~mJy for the C- and B-configuration datasets, respectively. This suggests that even at these high frequencies the jet-emission is still significant and there is no evidence for time variability in the continuum emission. 

The integrated line profile of the CO (1--0) emission, as observed with the VLA in C-configuration, is well represented by a Gaussian, although there is evidence of asymmetry in the line that is likely due to different velocity components being magnified differently (see Fig.~\ref{Fig:0751-spectrum}). In fact, a two-component Gaussian fit finds a redshifted Gaussian at a centroid of $283\pm25$ km~s$^{-1}$, a FWHM of $64\pm30$ km~s$^{-1}$ and a peak of $0.9\pm0.2$ mJy.  Nevertheless, as this fit has a similar reduced $\chi^2$ value to that of the single Gaussian fit, we cannot consider this asymmetry statistically significant. 
Therefore, for deriving the CO (1--0) line luminosity we use the single Gaussian fit parameters (peak of $1.5\pm 0.1$~mJy and a FWHM of $378\pm33$~km\,s$^{-1}$). The CO (1--0) line luminosity of MG J0751+2716 is $\mu_{\rm CO}\times L^\prime_{\rm CO (1-0)} = (25.7\pm5.1) \times 10^{10}$~K~km\,s$^{-1}$~pc$^2$ by following \citet{SolomonReview2005}, where $\mu_{\rm CO}$ is the magnification factor (see Table \ref{Tab:CO-parameters}). These observed line properties are in excellent agreement with the results obtained by \citet{Riechers2011}, who used the GBT and the VLA in D-configuration, and demonstrate that our VLA C-configuration dataset is recovering all of the CO (1--0) molecular gas associated with the lensed AGN. 

Our VLA C-configuration observations are about 3 times better in angular resolution (equivalent to an order of magnitude improvement in beam area) than those carried out previously and show that the CO (1--0) molecular gas emission from MG~J0751+2716 is smoothly extended in a north-east to south-west direction. Also, the peak of the integrated CO (1--0) intensity is offset with respect to the peak of the radio continuum emission, at a matched resolution, which demonstrates that the gas and the radio jets are not co-spatial (see Fig.~\ref{Fig:0751-moments}). The extended nature of the molecular gas distribution is also seen in velocity space, where we detect a velocity gradient that passes through the lens system. We note that the dominant velocity component is blue-shifted with respect to the systemic velocity of the AGN, which is again consistent with differential magnification effecting the line profile shape. \citet{Alloin2007} also detected a velocity gradient in their 0.5~arcsec resolved CO (3--2) emission with the Plateau de Bure Interferometer (PdBI), although there, the gradient is in a north-south direction and the CO (3--2) emission is contained within the Einstein radius. We also see a slight north-south velocity gradient in the blue-shifted component of our velocity map, but the beam-smearing makes this less pronounced (see Fig.~\ref{Fig:0751-moments}).

Finally, we note that there is no significant detection of the CO (1--0) molecular gas with the VLA in B-configuration, even though these data are at a similar depth to the C-configuration observations (see Fig.~\ref{Fig:0751-moments} first row). This suggests that for MG~J0751+2716, the molecular gas distribution is quite diffuse and resolved out at 0.2~arcsec-scales.

\subsection{JVAS~B1938+666} 

\begin{figure*}
\centering
	\includegraphics[width = 0.48\textwidth]{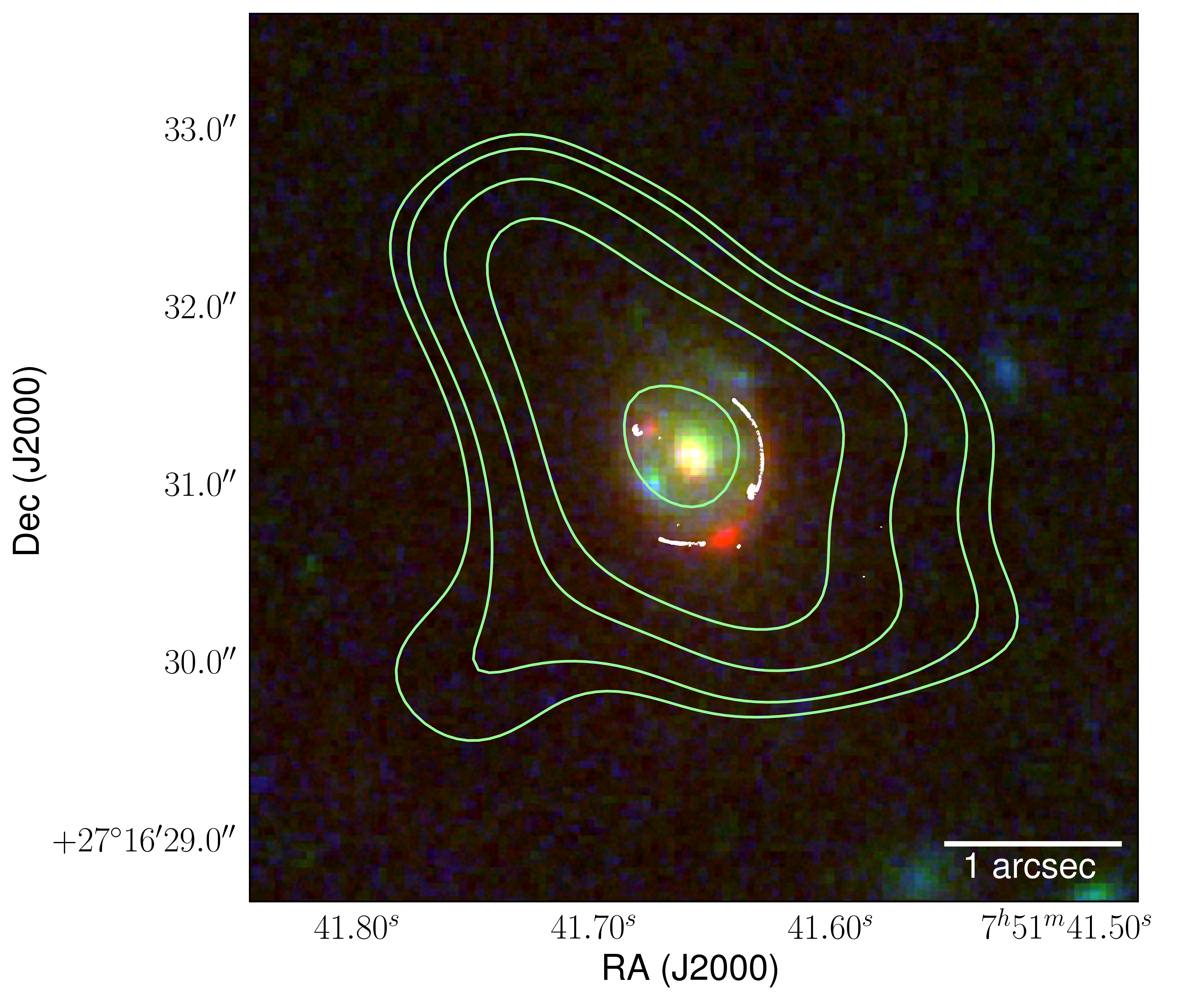}
	\includegraphics[width = 0.48\textwidth]{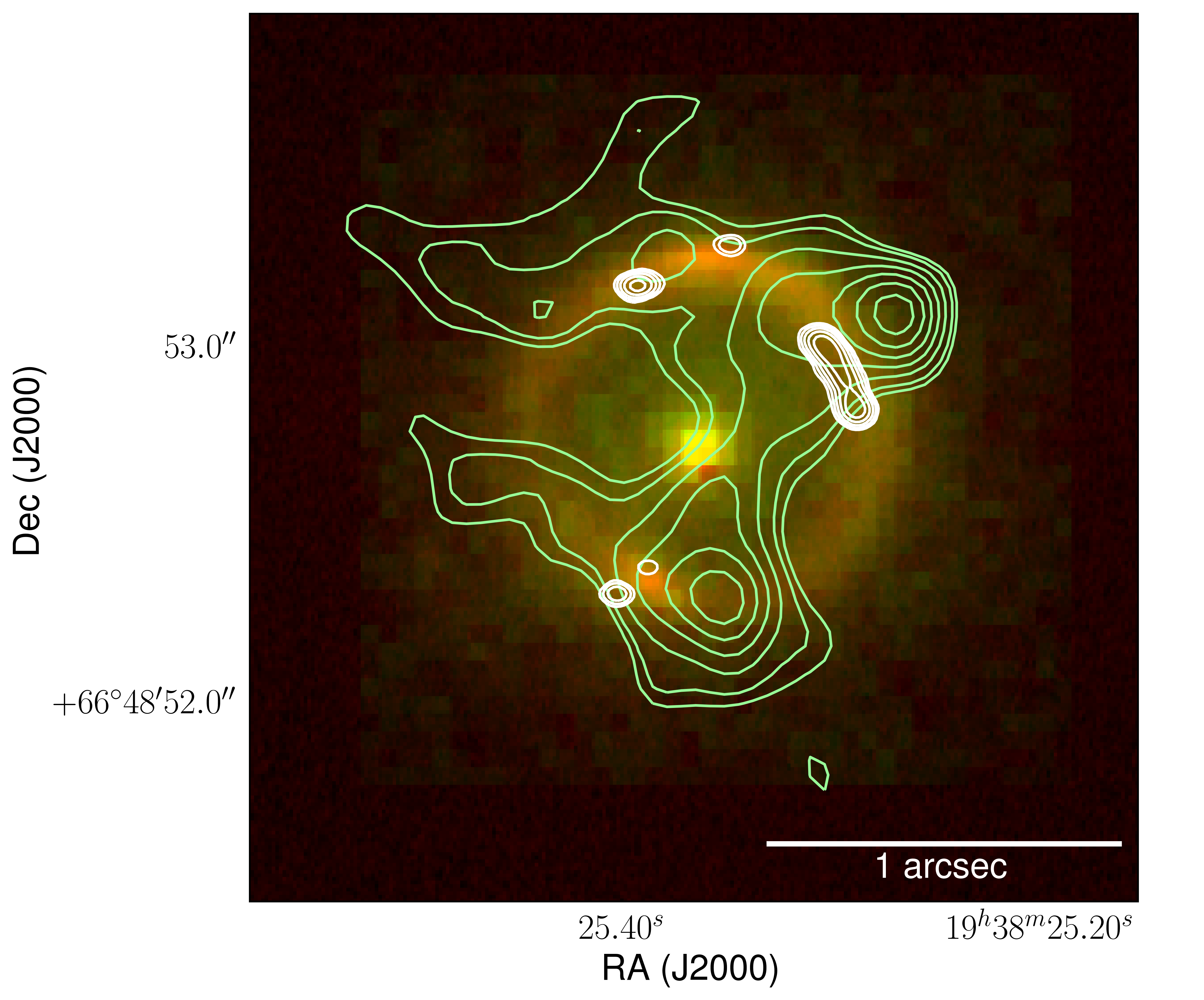}
    \caption{The lens-plane composite multi-wavelength images of MG~J0751+2716 (left) and JVAS~B1938+666 (right). In the case of MG~J0751+2716, a RGB rendering is made using the 2.12~$\mu$m Keck adaptive optics (red), the {\it HST} WFPC2-F814W (green) and the {\it HST} WFPC2-F555W (blue), and the white contours show the radio continuum emission from global VLBI imaging at 1.7 GHz \citep{Spingola2018}. In the case of JVAS~B1938+666, a RGB rendering is made using the 2.12~$\mu$m Keck adaptive optics (red), the {\it HST} NICMOS-F160W (green) and the {\it HST} WFPC2-F814W (blue), and the white contours show the radio continuum emission from VLA A-configuration imaging at 37.7 GHz. The green contours show the CO (1-0) moment zero emission for both lensed objects, and increase by factors of three. The first contour level is at 43 and 30~mJy beam$^{-1}$ for MG~J0751+2716 and JVAS~B1938+666, respectively. The different wavebands have been aligned using the lens model to fix the position of the lensing galaxy with respect to the lensed emission.}
    \label{Fig:RGB}
\end{figure*}

The optical and NIR imaging of JVAS~B1938+666 clearly resolves the early-type foreground lensing galaxy and the almost complete Einstein ring of the AGN host galaxy (see Fig.~\ref{Fig:1938-hst-keck}), as was previously reported by \citet{King1997} and \citet{Lagattuta2012}. We find that the radio continuum emission at 37.7 GHz is not co-incident with the AGN host galaxy, but is extended beyond the stellar halo (see Fig.~\ref{Fig:RGB}): the offset between the radio and NIR emission is significant considering the astrometric precision of the Keck  adaptive optics (4~mas) and the VLA A-array observations (30~mas; see below). Both the doubly and quadruply imaged parts and the extended gravitational arc that were found at lower frequencies \citep{King1997} are also detected in our B- and A-configuration datasets, with measured continuum flux-densities of $S_{\rm 37.7~GHz} = 59.1\pm3.0$ mJy and $S_{\rm 37.7~GHz} = 61.4\pm3.0$ mJy, respectively.  Again, the lack of strong variability and the offset in the position of the radio components from the AGN host galaxy are both consistent with non-thermal jet-emission. This interpretation is also in agreement with the parametric lens modelling of the NIR and 5 GHz radio data by \citet{King1997}. In addition, unlike in the case of MG~J0751+2716, we find no evidence for the radio core in the high-frequency data, which we would expect to be coincident with the centre of the AGN host galaxy. 

The CO (1--0) emission line has a more complicated profile than can be modelled by a single Gaussian function, having an asymmetric structure with one strong peak centred at $-190$~km\,s$^{-1}$ and another fainter peak centred at $+180$~km\,s$^{-1}$ relative to the systemic velocity of the AGN; this kind of line profile is consistent with a differential magnification of a double horn profile from a rotating gas disc (e.g. \citealt{Popovic2005,Banik2015,Rybak2015b,Leung2017,Paraficz2018, Stacey2018b}). A Gaussian fit to the profile gives a line peak of $0.9\pm 0.4$~mJy, a FWHM of $683\pm280$~km\,s$^{-1}$, and a CO (1--0) line luminosity of $\mu_{\rm CO}\times L^\prime_{\rm CO} =( 20.2 \pm 4.0) \times 10^{10}$~K~km\,s$^{-1}$~pc$^2$ (see Table \ref{Tab:CO-parameters}). The overall line structure and FWHM are consistent within the large uncertainties with the results of \citet{Sharon2016}, who first detected CO (1--0) from JVAS~B1938+666 using the VLA in D-configuration,  demonstrating that our higher angular-resolution imaging is recovering most, if not all, of the CO (1--0) molecular gas emission.

Our B-configuration spectral-line imaging of the CO (1--0) has an angular resolution that is almost a factor of 10 better than previous studies. These data fully resolve the molecular gas from JVAS~B1938+666 into an Einstein ring that closely matches the position of the AGN host galaxy emission, but is also clearly more extended. With a beam-size of $\sim0.25$~arcsec, this represents the highest angular resolution imaging of extended CO (1--0) emission from a high redshift object, which is further enhanced in angular resolution by the gravitational lensing. Overall, the surface brightness distribution of the molecular gas is rather smooth when tapering the visibilities, with some higher surface brightness clumps, and a clear highest peak brightness component to the north-west of the AGN host galaxy and the non-thermal radio emission. We highlight that the smoothness of the CO (1--0) intensity might be due to the limited angular resolution and the tapering of the visibilities. We find that the Einstein ring of the molecular gas has significant velocity structure, with the most red-shifted component being in the doubly imaged region and associated with the brightness emission, and the blue-shifted component being quadruply imaged and forming the Einstein ring; this surface brightness distribution of the gas explains the asymmetric line profile described above. Also, the most blue-shifted component extends to the east of the molecular gas ring and the AGN host galaxy. This can be interpreted as a velocity gradient along the east-west direction across the lens system in the moment one map.

\begin{table*}
\caption{Observed line and continuum parameters for the VLA observations of MG~J0751+2716 and JVAS~B1938+666. (Left to right) Given is the target name, the VLA configuration used during the observations, the continuum flux-density, the CO (1--0) spectral line peak, FWHM and integrated line intensity (all three from a Gaussian fit), the specific CO line luminosity  (not corrected for the lensing magnification of the CO $\mu_{\rm CO}$) that is calculated following \citet{SolomonReview2005}, the CO magnification factor and the major and minor axes of the two-dimentional Gaussian fit to the CO emitting region in the source plane (see Sections ~\ref{Sec:results_1} and ~\ref{Sec:results_2}).}

	\begin{tabular}{lllllllcc}    
	\hline
	Target 			& Config.	& $S_{\nu}$	& $\Delta v_{\rm FWHM}$ 	& $I_{\rm CO}$  	& $\mu_{\rm CO} \times L^\prime_{\rm CO}$ &   $\mu_{\rm CO}$ & Major axis & Minor axis\\
	 				& 					&  (mJy)		&  (km\,s$^{-1}$) 			& (Jy~km\,s$^{-1}$) 	& ($10^{10}$ K~km\,s$^{-1}$~pc$^2$) & & (kpc) & (kpc)\\
	\hline
	MG~J0751+2716 & C & $1.5 \pm 0.1$ & $378\pm33$ & $0.56 \pm 0.12$ & $25.7\pm 5.1$ & $2.1\pm0.8$ & $20\pm3$ & $6\pm2$ \\
	\hline
	JVAS~B1938+666 	& B					& $0.9\pm 0.4$	& $683\pm280$		 & $0.94 \pm 0.23$  & $20.2 \pm 4.0$ &  $8.7\pm3.1$ & $5 \pm 2$ & $2\pm 1$\\
    \hline
	\end{tabular}
	
    \label{Tab:CO-parameters} 
\end{table*}

\section{Intrinsic source properties}
\label{sec:results} 

We now describe the multi-wavelength source reconstruction of MG J0751+2716 and JVAS B1938+666, shown in Figs.~\ref{Fig:source_0751} and \ref{Fig:source_1938}, respectively, in order to infer the intrinsic properties of the two lensed objects. 

\subsection{Lens modelling}
\label{sec:lensmodelling}

In order to obtain the source plane surface brightness distribution of both targets for the optical and NIR data we use the pixellated source reconstruction technique developed by \citet{Vegetti2009} and further improved by \citet{Ritondale2019} and \citet{Rizzo2018}. This code models the lens potential, the surface brightness distribution of the foreground lensing galaxy and the surface brightness distribution of the background source simultaneously in a fully Bayesian framework. We find the light profile of the lensing galaxy is well described by two elliptical S\'ersic profiles for MG~J0751+2716 and one elliptical S\'ersic profile for JVAS~B1938+666 (consistent with \citealt{Lagattuta2012} for the latter).

Crucial to carrying out a multi-wavelength analysis is the correct alignment of the different datasets, which have uncertainties in the absolute astrometry of up to $\sim0.3$~arcsec. We align the optical, NIR and radio data using the inferred positions of the lensing mass, which provides a common reference point between all datasets. We use the best quality lens mass models from \citet{Lagattuta2012} and \citet{Spingola2018}, keeping all of the lens mass model parameters fixed except for the lensing mass position, which is allowed to vary, and apply a backward ray-tracing approach to determine the maximum likelihood source whilst optimising for the position of the lensing mass. This approach provides a precise measurement of the relative astrometry between the datasets, and is another advantage of multi-wavelength studies that use gravitationally lensed objects. The uncertainty in the lensing mass position, which translates to the uncertainty in the multi-wavelength alignment, is determined using {\sc mulitnest} \citep{Feroz2009} to sample the posterior density distribution. 
The uncertainty on the position of the lensing mass of MG~J0751+2716 is inferred at 2.5~mas ($1\sigma$) precision from the VLA dataset and at 1.5~mas precision from the Keck adaptive optics dataset. For JVAS~B1938+666, the precision on the lensing mass position is 30 mas from the VLA dataset and 4~mas from the Keck adaptive optics dataset.

In order to reconstruct the radio continuum and spectral line emission in the source plane, we use a Bayesian pixellated lens modelling code that fits directly in the visibility space. This new extension of the \citet{Vegetti2009} method will be presented in detail in a follow-up paper (Powell et al. in prep.). Due to the low signal-to-noise ratio detection of the CO (1--0) line, we do not use the full spectral resolution of the available line channels, but instead we average the data in frequency.

For MG~J0751+2716, we cover the velocity structure of the line emission with 12 channels, each with a velocity width of 43.7 km\,s$^{-1}$. Moreover, we use only baselines shorter than 1410 m, where most of the signal is detected. This $uv$-cut leads to an angular resolution of about 1.6 arcsec. For JVAS~B1938+666 we cover the velocity structure with six channels with a velocity width of 63.6 km\,s$^{-1}$ and we apply a $uv$-cut of 4200 m, which corresponds to an angular resolution of about 0.4 arcsec. For both targets we compute the moment maps in the standard manner, by integrating the de-lensed spectral channels along the velocity axis to obtain the total intensity map and intensity-weighted velocity field\footnote{We follow the methodology outlined at \url{https://casa.nrao.edu/docs/casaref/image.moments.html}.}, and we include only pixels that have a signal-to-noise ratio larger than 5.

To parameterize the properties of the lensed AGN host galaxies, we fit S\'ersic profiles to the reconstructed optical/NIR emission using {\sc galfit} in the source plane \citep{Peng2010} to determine the effective radius ($R_{\rm e}$) and the S\'ersic index ($n$) of their light distribution. For the CO (1--0) emission, we fit a two-dimensional Gaussian to estimate the major and minor axis of the molecular gas distribution. We also fit the CO (1--0) emission with a S\'ersic profile to obtain the $R_{\rm e}$, which will be used to estimate the dynamical mass. To calculate the intrinsic luminosities, we sum the emission in the reconstructed source plane, as this removes the effects of differential magnification from our analysis.


\subsection{Source-plane morphology of MG~J0751+2716}
\label{Sec:results_1}

\begin{figure*}
\centering
	\includegraphics[width = 1\textwidth]{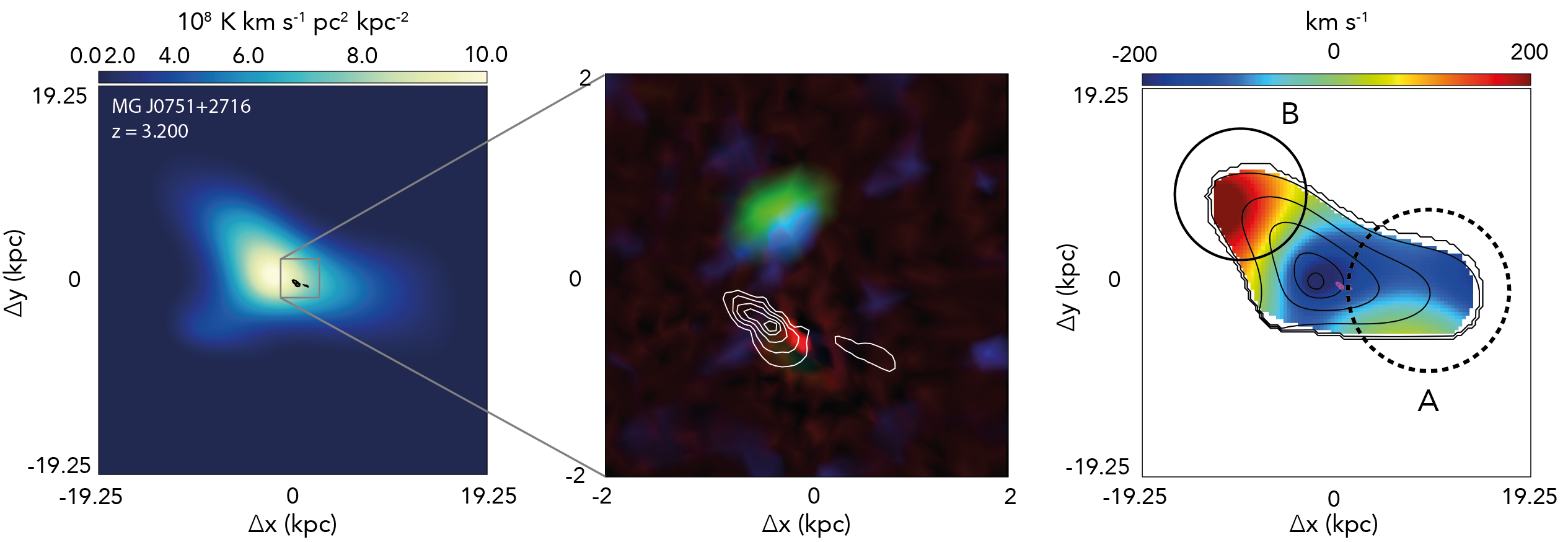}
    \caption{Composite source-plane reconstruction of MG~J0751+2716. (Left) The intrinsic CO (1--0) surface density distribution (colour map), where the black contours indicate the non-thermal radio emission, which are set to 20, 40, 60, 80 and 95 per cent of the peak intensity. (Centre) A zoom of the central part of the system, where the rest-frame UV/optical emission is shown in red (Keck-AO), green (F814W) and blue (F555W), and the white contours indicate the radio emission, which are set to 20, 40, 60, 80 and 95 per cent of the peak intensity. (Right) The intrinsic velocity field of MG~J0751+2716. The black contours trace the moment zero emission, which are set to 1, 20, 40, 60, 80 and 95 per cent of the moment zero peak. The velocity gradient seen in the north-south direction is indicated by the dashed circle (region A), while the north-east component at average velocity $+$90 km\,s$^{-1}$ is indicated by the solid circle (region B). The purple contours indicate the non-thermal radio emission, which are set to 20, 40, 60, 80 and 95 per cent of the peak intensity.}
    \label{Fig:source_0751}
\end{figure*}

The multi-wavelength source-plane reconstruction of MG~J0751+2716 is shown in Fig.~\ref{Fig:source_0751}. The object has a complex morphology, where the two distinct rest-frame optical components are found to be separated by $\sim 1$~kpc in projection, the jet-emission from the AGN is extended by $\sim 2$~kpc in projection, and the diffuse molecular gas distribution extends far beyond the optical and radio components, with an effective radius of $17\pm5$~kpc. We highlight that the optical-radio continuum and CO-radio continuum offsets are also detected in the lens plane at high significance ($\sim200$ and $\sim300$~mas, respectively), given the uncertainty on the lensing galaxy position, which is of order 2~mas (see Sec.~\ref{sec:lensmodelling}).

When fitting the gas distribution with a two dimensional Gaussian, we found that the CO (1--0) emission extends with a major axis of $20 \pm 3$~kpc and a minor axis of $6 \pm 2$~kpc in projection. Nevertheless, we believe that these sizes should be considered as upper limits. Given the limited angular resolution of our observations, MG~J0751+2716 is barely spatially resolved (covered only by $\sim3$ beam elements in the lens plane ( see Fig.~\ref{Fig:0751-moments}). As our  B-configuration observations at higher angular resolution did not detect any emission from the CO (1--0), this might be taken  as evidence for the gas being intrinsically extended. However, the non-detection could also be related to the surface brightness sensitivity of the B-configuration observations,  which would need to be about 10 times deeper to detect all of the resolved emission from the CO (1--0) found with the VLA in C-configuration.  For this reason, we cannot rule out that the cold molecular gas contains several compact structures in addition to the smooth extended component we detect here. Given the extremely deep imaging with the VLA that would be required to test whether there are any compact features in the CO (1--0) emission, imaging of higher excitation CO (3--2) with ALMA will likely provide a better estimate of the structure of the CO gas in this object.

The peak integrated intensity of the molecular gas is offset with respect to the radio and optical emission by almost 2 kpc in projection.  Moreover, the source-reconstructed velocity field shows a gradient that is perpendicular to the major axis of the gas intensity distribution (called region A in Fig.~\ref{Fig:source_0751}), which is consistent with the CO (3--2) velocity field measured by \citet{Alloin2007}. This velocity gradient could be interpreted as a rotating disc in the north-south direction \citep[e.g.][]{Smit2018}, but it could also be due to the two optical components moving at different velocities, as for example, in the case of interacting galaxies \citep[e.g.][]{Jones2010}. In addition, there is a component with an average velocity of around $+90$~km\,s$^{-1}$, which is located in the north-east part of the velocity field, and is offset with respect to the bulk of the CO (1--0) emission associated with the AGN by $\sim10$~kpc in projection (indicated as region B in Fig.~\ref{Fig:source_0751}). The intrinsic line luminosity is found to be  $L^\prime_{\rm CO} = (1.6\pm0.6) \times 10^{11}$ K km\,s$^{-1}$ pc$^2$, which is in agreement with the $L^\prime_{\rm CO} = (2.2\pm0.2) \times 10^{11}$ K km\,s$^{-1}$ pc$^2$ determined by \citet{Alloin2007}, who used the CO (3--2) emission. 

For the blue and red optical components, we find effective radii of $R_{\rm{e}}^{\rm blue} = 530 \pm 90$~pc and $R_{\rm{e}}^{\rm red} =300 \pm 40$~pc, respectively. The compact size of the red component is robust given the excellent sensitivity and angular resolution of the Keck adaptive optics observations. Instead, the blue component shows evidence for a more extended emission in the \textsl{HST}--WFPC2/F555W observations, but the signal-to-noise ratio is too low to confidently recover the possible optical extended arcs in the source plane (recall that the Bayesian lens modelling procedure derives the best source model given the data). The S\'ersic index of the blue component is $n=0.9\pm0.3$, which indicates a more disky structure, while the red component is closer to a De Vaucouleurs profile as $n= 4.5\pm0.2$.

The radio jets are aligned with the major axis of the red component and are centred on it, but extend beyond the rest-frame optical emission of the host galaxy (see Fig.~\ref{Fig:source_0751}). The angular extent of the radio jets of just $\sim 2$~kpc may be an indication for the AGN being young or for the presence of a dense surrounding medium (or partly due to projection effects). We note that the eastern jet is also in the direction of the peak in the CO molecular gas distribution, which could be evidence of a jet-gas interaction in the dense intra-cluster medium, but further higher resolution data would be needed to confirm this. 

If we assume that the two optical components are both galaxies, the source morphology of MG~J0751+2716 could be interpreted as a possible early-stage merger between the two, which are embedded in an extended molecular gas reservoir. Also, the merger scenario between these two galaxies can provide a suitable explanation for the high FIR luminosity observed with \textsl{Herschel} ($\mu_{\rm FIR} \times L_{\rm{FIR}} = 10^{13.4}$~M$_{\odot}$; \citealt{Stacey2018}), as starbursts are thought to be triggered through galaxy interactions \citep[e.g.][]{Sanders1988, Genzel2001, Hopkins2006, Hopkins2008}. 

Alternatively, it could be that the blue component is a Ly$\alpha$ cloud illuminated by the AGN continuuum, as the \textsl{HST} F555W filter covers the wavelength range of the Ly$\alpha$ emission at $z = 3.2$. These Ly$\alpha$ clouds are often found associated with radio AGN, and they are believed to consist of re-scattered Ly$\alpha$ photons by the neutral hydrogen surrounding the continuum source \citep[e.g][]{Hayes2011, North2017}.

\subsection{Source-plane morphology of JVAS~B1938+666}
\label{Sec:results_2}

\begin{figure*}
\centering
	\includegraphics[width = 1.0\textwidth]{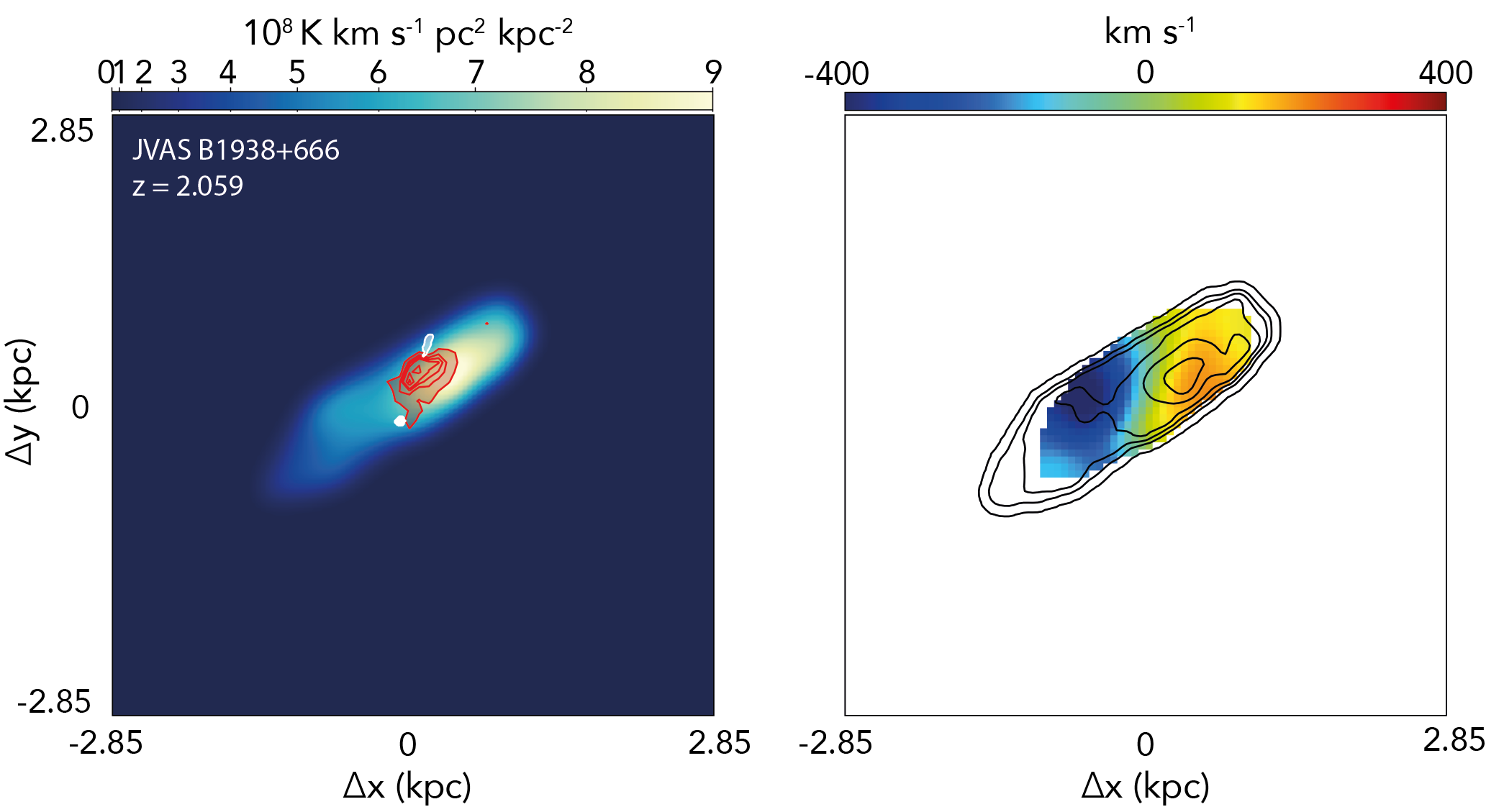}
    \caption{Composite source-plane reconstruction of JVAS~B1938+666. (Left) The colour map is the CO (1--0) surface density distribution, while the red and white contours indicate the rest-frame optical (Keck-AO) and radio emission, respectively, where the contour levels are set to 20, 40, 60, 80 and 95 per cent of the peak optical and radio intensity. (Right) The intrinsic velocity field of JVAS~B1938+666, where the black contours trace the CO (1--0) moment zero emission, which are set to 1, 20, 40, 60, 80 and 95 per cent of the moment zero peak.}
    \label{Fig:source_1938}
\end{figure*}

The multi-wavelength reconstruction of the JVAS~B1938+666 source-plane is shown in Fig.~\ref{Fig:source_1938}. The NIR emission consists of a single component with an effective radius of $R_{\rm{e}} =460 \pm 70$~pc and a S\'ersic index of $n= 1.5\pm0.2$.
This compact red galaxy is likely hosting the AGN, which is composed of two almost unresolved lobes/hotspots that are separated by $\sim 300$~pc in projection. Given the flat radio spectra of the two components and their symmetric emission around a plausible dust-obscured core, it has been suggested that JVAS~B1938+666 is likely a compact symmetric object (CSO; \citealt{King1997}). Similar to MG J0751+2716, it is not clear if the radio source is young, or whether the radio source is contained within the medium around the host galaxy.

A two dimensional Gaussian fit finds that the de-lensed integrated line intensity of the CO (1--0) extends with a major axis of $5 \pm 2$~kpc and a minor axis of $2 \pm 1$~kpc in projection over the entire system, perpendicularly to the compact radio emission. A S\'ersic fit to the molecular gas distribution gives an effective radius of $R_e = 1.5\pm0.5$ kpc. The intrinsic line luminosity is $L^\prime_{\rm CO} = (2.5 \pm 0.8) \times 10^{10}$ K km s$^{-1}$ pc$^2$. Our lensing-corrected moment zero map is dominated by the redshifted component of the molecular gas distribution, and its peak is offset by almost $1$~kpc in projection with respect to the radio and the optical emission (see Fig.~\ref{Fig:source_1938}). This can also be seen as a decrease in the surface brightness of the CO (1--0) where the AGN emission and host galaxy are located, which may indicate a decrement in the molecular gas at low excitation in the regions closer to the AGN. This effect has also been seen in other low $J$-level molecular gas distributions around lensed AGN when observed at sufficiently high angular resolution (e.g. \citealt{Paraficz2018}). The velocity distribution shows a clear gradient across the major axis, which can be interpreted as multiple velocity source components, or more likely, as a rotating molecular  gas disc. The inclination-corrected maximum rotational velocity of the gas, in this latter scenario, is $v_{\rm max}= 355\pm150$~km\,s$^{-1}$, which has been estimated by measuring the maximum velocity along the major axis in the reconstructed source. This value is much larger than the typical maximum velocities of low-redshift spiral galaxies \citep[e.g.][]{Frank2016}, but similar to those of gravitationally lensed disc galaxies hosting an AGN (e.g. \citealt{Venturini2004,Paraficz2018}). However, we highlight that the estimate $v_{\rm max}$ is likely affected by beam smearing \citep[e.g.][]{DiTeodoro2015}. Nevertheless, the low signal-to-noise ratio of the line emission prevents us to robustly model the source data cube using sophisticated kinematical modelling tools that take into account also the gravitational lensing effect \citep{Rizzo2018}.

Overall, JVAS~B1938+666 has a somewhat similar source-plane morphology to MG~J0751+2716; there is a compact red galaxy that hosts a dust obscured AGN with small radio jets that can be seen just beyond the stellar component of the galaxy. The molecular gas distribution can be interpreted as a possible disc, but unlike in the case of MG J0751+2716, the extent of the molecular gas is much smaller, more regular and has a higher surface brightness.

\section{Resolved molecular gas emission at high redshift}
\label{Sec:results_3}

In this section, we present the results from our analysis of the spatially resolved CO (1--0) observations of MG J0751+2716 and JVAS~B1938+666. Our aim is to determine the physical properties of the cold molecular gas within these two star-forming/AGN composite galaxies at $z\simeq2$--3.

\subsection{CO--H$_2$ conversion factor and molecular gas mass}
\label{Sec:alpha_CO}

To determine the molecular gas mass ($M_{\rm gas}$) from our observations requires some knowledge of the conversion factor ($\alpha_{\rm CO}$) between H$_2$ and CO,
\begin{equation}
\alpha_{\rm CO} = \frac{M_{\rm gas}}{L^\prime_{\rm CO}}.
\end{equation}
Often, two values for the conversion factor $\alpha_{\rm CO}$ are adopted in literature, according to the galaxy type. For quiescent late-type galaxies, such as our own Milky Way, the most common value is $\alpha_{\rm CO}=4.3$~M$_{\odot}$ (K km s$^{-1}$ pc$^2$)$^{-1}$ \citep{Bolatto2013}, while in the case of high redshift starburst galaxies, $\alpha_{\rm CO}=0.8$~M$_{\odot}$ (K km\,s$^{-1}$ pc$^2$)$^{-1}$ has been usually assumed \citep{Aravena2016}. However, observational and theoretically motivated studies have both shown that using $\alpha_{\rm CO}=0.8$ M$_{\odot}$~(K km\,s$^{-1}$ pc$^2$)$^{-1}$ may not be correct for converting the observed line luminosity into the molecular gas mass for high-redshift galaxies \citep{Narayanan2012a, Hodge2012}.  For instance, this value for the CO--H$_2$ conversion factor leads to gas fractions of $f_{\rm gas}\sim80$ per cent, which is much higher compared to what is predicted from simulations ($f_{\rm gas} \sim 30$ per cent). Moreover, independent estimates of $\alpha_{\rm CO}$ using different methods find that the CO--H$_2$ conversion factor covers a large range of values at high redshift \citep[e.g.][]{Daddi2010, Ivison2011, Hodge2012}, indicating that its value cannot be universally assumed, but must be estimated for each individual case. Alternatively, the conversion factor for lensed galaxies at high-redshift can be estimated by comparing the enclosed dynamical mass and the luminosity-derived molecular gas mass (e.g. \citealt{Paraficz2018}). However, this approach requires knowledge of the stellar mass and dark matter contributions to the total enclosed mass, which are not well constrained by our current datasets for MG~J0751+2716 and JVAS~B1938+666.

\citet{Narayanan2012b} recently demonstrated that $\alpha_{\rm CO}$ requires a complex description, and that there is a wide continuum spectrum of values for the CO--H$_2$ conversion factor depending on several physical parameters. They developed a functional form that relates the CO surface brightness and the metallicity of the specific phase of the gas within a galaxy to $\alpha_{\rm CO}$, such that, 
\begin{equation}
\alpha_{\rm CO} = \frac{10.7 \times \langle W_{\rm CO}\rangle^{-0.32} }{Z^{\prime0.65}} \, ,
\label{eqn:gas}
\end{equation}
where $\langle W_{\rm CO}\rangle$ is the CO surface brightness for a uniformly distributed molecular gas luminosity (K km\,s$^{-1}$), and $Z^\prime$ is the metallicity of the gas phase \citep{Narayanan2012b}. We approximate $\langle W_{\rm CO}\rangle$ as $L^\prime_{\rm CO}$/area, where the area is given by the two-dimensional Gaussian fit in the source plane (see Sections \ref{Sec:results_1} and \ref{Sec:results_2}). Since we do not have measurements of the gas-phase metallicity in the cases of MG~J0751+2716 and JVAS~B1938+666, we assume solar metallicity, which is a typical value for galaxies at $z\sim 2$ \citep[e.g.][]{Erb2006}. We note that the CO surface brightness in our two lensed AGN is not homogeneously distributed (see Figs.~\ref{Fig:source_0751} and \ref{Fig:source_1938}). As a consequence, our values for $\alpha_{\rm CO}$ will be upper limits. 

From Eqn. \ref{eqn:gas}, we find individual CO--H$_2$ conversion factors of $1.5 \pm 0.5$ and  $1.4 \pm 0.3$~M$_{\odot}$ (K km\,s$^{-1}$ pc$^2$)$^{-1}$ for MG J0751+2716 and JVAS~B1938+666, respectively.  This results in molecular gas masses of $M_{\rm gas} = (2.5 \pm 0.8) \times 10^{11}$~M$_{\odot}$ and $M_{\rm gas} = (3.4 \pm 0.8)\times 10^{10 }$~M$_{\odot}$ for MG~J0751+2716 and JVAS~B1938+666, respectively. These masses are larger than those of local ultra-luminous infrared galaxies (ULIRGs; \citealt{Papadopoulos2012}), but consistent with distant ULIRGs and AGN-host galaxies, which are suggested to trace the most massive galaxies at high redshifts \citep{Chapman2009, Daddi2010, Braun2011, Casey2011, Aravena2013, Tacconi2013, Ivison2013, Sharon2016, Noble2018}.

\subsection{Dynamical masses}
\label{Sec:dynamical_mass}

From our resolved CO (1--0) velocity maps of MG J0751+2716 and JVAS B1938+666, we can estimate the enclosed dynamical mass within the effective radius $R_e$, under the assumption that these velocity fields are consistent with a rotating disc and that the systems are virialized. As discussed above, in the case of JVAS~B1938+666, the velocity field is well-ordered and shows convincing evidence for rotation; the systemic velocity of the molecular gas is centred on the host galaxy, and the red and blue-shifted components form an elongated and symmetric structure. There is also no evidence of companion galaxies associated with the AGN host galaxy. The case of MG J0751+2716 is less clear, as the molecular gas has a more complex morphology, is much more extended and there is evidence of multiple optical components within the system.

We estimate the enclosed dynamical mass (in M$_{\odot}$) as
\begin{equation}
M_{\rm dyn} = \frac{[{\rm v}_e/\sin(i)]^2 R_e}{G},
\end{equation}
where $R_e$ is the effective radius of the molecular gas disc as measured by the Sersic fit (see Sections \ref{Sec:results_1} and  \ref{Sec:results_2}), $i$ is the inclination angle, and ${\rm v}_e$ is the velocity at $R_e$ of the CO (1--0) line. To have an approximate estimate of the inclination angle, we use the ratio between the observed minor and major axis of the gas emission, as determined above. We find that the dynamical mass in MG~J0751+2716 is $M_{\rm dyn}(<R_e$) $= (4.2 \pm 1.6) \times 10^{11}$ M$_{\odot}$, where the effective radius is $R_e= 17\pm 5$ kpc, the inclination angle is $i=72\pm20$~deg and ${\rm v}_e = 200\pm50$~km\,s$^{-1}$. For JVAS~B1938+666 the mass is estimated to be $M_{\rm dyn}(<R_e$) $= (5.8 \pm 0.5) \times 10^{10}$ M$_{\odot}$, where $R_e= 1.5\pm0.5$ kpc, ${\rm v}_e = 350\pm70$~km\,s$^{-1}$ and the inclination angle is $i=67\pm15$~deg. 

These values are larger than the molecular gas mass estimates using the CO luminosity, which is as expected. Nevertheless, they are comparable to the dynamical masses of starburst galaxies at similar redshifts (e.g. \citealt{Hodge2012}). However, since there is morphological evidence that at least one of our lensed AGN is maybe going through a merger phase, our dynamical mass estimate is only indicative.

\subsection{Gas fractions within dust obscured AGN}
\label{Sec:gas_fraction}

\begin{figure}
\centering
	\includegraphics[width = 0.48\textwidth]{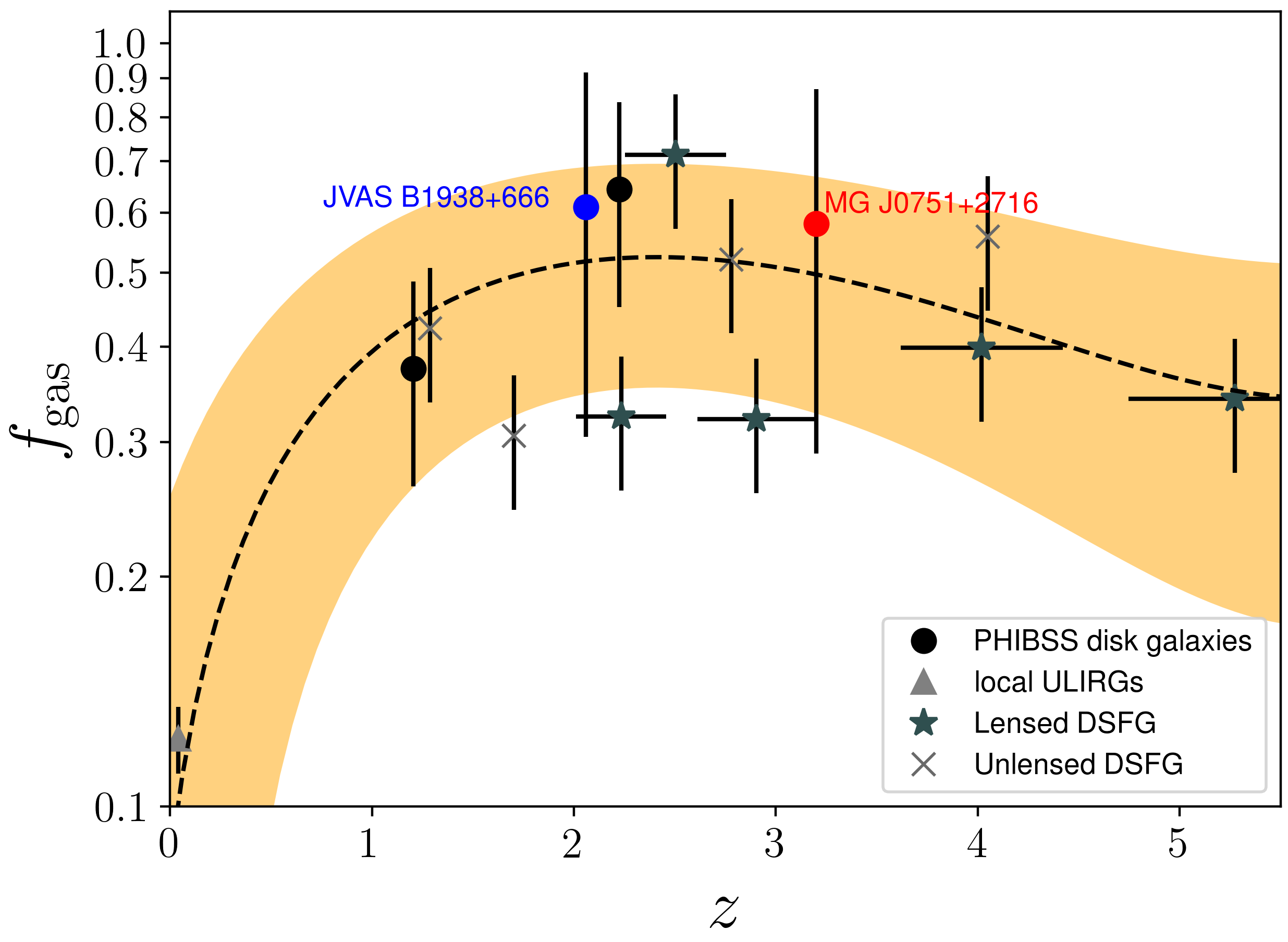}
    \caption{Comparison between the gas fraction in MG J0751+2716 and JVAS B1938+666 (red and blue filled circles) with different galaxy populations as a function of redshift. The gas fraction is estimated as $f_{\rm gas} = M_{\rm gas} / M_{\rm dyn}$. The literature values are from \citet{Tacconi2013} (Plateau de Bure HIgh-z Blue Sequence Survey, PHIBBS galaxies), \citet{Downes1998} (low-$z$ ULIRGs), \citet{Harris2012} (gravitationally lensed DSFG) and \citet{Bothwell2013} (unlensed DSFG). The dashed line represents the best fit, while the shaded yellow area is the formal uncertainty on the fit, which is consistent with an increasing $f_{\rm gas}$ until redshift $z=2$ and constant $f_{\rm gas}$ after $z=2$.} 
    \label{Fig:gas_frac}
\end{figure}

With our estimate of the molecular gas and dynamical masses, we can now calculate the gas fraction within the two lensed AGN. This typically also requires knowledge of the stellar mass component, which cannot be robustly estimated for our two de-lensed targets. However, if we assume that their ISM is molecular dominated, we can approximate the gas fraction to,
\begin{equation}
f_{\rm gas} = M_{\rm gas} / M_{\rm dyn}.
\end{equation}
We find gas fractions of $f_{\rm gas} = 0.61\pm0.20$ and $f_{\rm gas}= 0.58\pm0.15$ for JVAS B1938+666 and MG~J0751+2716, respectively.
Our findings of large gas fractions in these two galaxies at $z>2$ can be taken as an evidence for the importance of gas-rich mergers at high redshift. In fact, our $f_{\rm gas}$ are in agreement with the theoretical expectation that high redshift star-forming galaxies should be more gas-rich than those galaxies at redshift $z=0$ (see Fig.~\ref{Fig:gas_frac}; \citealt{Daddi2010, Tacconi2010, Lagos2011, Narayanan2012b, Bothwell2013, Aravena2016, Aravena2019}). This is because, according to the current galaxy evolution scenario, at those epochs the galaxy formation is dominated by gas-rich mergers, leading to dust and gas masses much larger than those of local galaxies \citep{Hopkins2009a}. Moreover, for our two lensed AGN, we find that the gas fractions are consistent within the uncertainties with those of dusty star-forming galaxies (DSFGs) at similar epochs, which coupled with their FIR dust properties \citep{Stacey2018}, would again add to the evidence that these objects are dust obscured AGN--starburst composites.

\section{Discussion}
\label{sec:discussion} 

We now discuss our results for the two radio-loud lensed AGN that were observed at high angular resolution as part of our multi-wavelength campaign. These targets were chosen because of their extended radio jets, large CO (1--0) luminosities and large lensing magnifications, and therefore should provide excellent sites to investigate the effect of mechanical feedback at high redshift. In the previous sections, we determined the intrinsic properties of the various components (stars, radio jets and molecular gas) after correcting for the gravitational lensing magnifications, which we now compare with other high redshift galaxies. We first discuss the properties of the molecular gas distributions, and then present an analysis of the optical properties, with a view of better understanding the effect AGN activity has on the star formation and build-up of the stellar population within AGN host galaxies.

\subsection{Evidence for extended molecular gas reservoirs}
\label{Sec:extended_gas_reservoir}

The standard model of galaxy formation predicts that the bulk of the cosmic stellar mass assembly occurred at early epochs ($z>2$), and on fast timescales through a process of dusty, gas-rich mergers \citep{Sanders1988,Hopkins2009a}. In this scenario, hydrodynamical simulations have shown that starburst galaxies with SFRs of $>200$~M$_{\odot}$~yr$^{-1}$ grow in stellar mass by 1.5 dex between redshifts 3 and 1 \citep{Vogelsberger2014}. Such extreme bursts of star formation are thought to be triggered through gas-rich mergers, which given the substantial obscuration due to dust, are typically studied at rest-frame FIR wavelengths due to the re-emission of the UV radiation from new stars \citep{Hodge2012}. As these galaxies continue to grow via mergers, a dust-obscured quasar phase is expected as the super massive black hole builds-up via gas accretion. At this point, radiative feedback from the intense UV radiation field from the quasar should heat the gas and dust, possibly limiting star formation in the vicinity of the AGN, and perhaps further throughout the galaxy (up to kpc scales), until the quasar is no longer obscured. During this, and possibly later active phases for the super massive black hole, it is expected that radio jets inject mechanical feedback into the system, clearing the galaxy of a significant fraction of the dust and gas to the point that star formation is halted. Without significant gas accretion, the stellar population of the galaxy will continue to evolve to present day until a passive galaxy is formed \citep{Sanders1988}.

Our multi-wavelength dataset for MG J0751+2716 and JVAS B1938+666 is somewhat consistent with this picture. These two objects are known to harbour an active central engine, given the powerful radio jets that originate from the host galaxy. Since there is no evidence of quasar-like morphology from optical imaging, either the black hole is heavily obscured by dust or the significant reddening is due to active star formation in the two obscured AGN. These conclusions are consistent with the red rest-frame optical colours of the host galaxies and the extreme levels of heated dust found by \citet{Stacey2018}. Therefore, these AGN are likely moving through an obscured quasar phase. Based on the large IR luminosities, and assuming that this heating is due predominantly to star formation, \citet{Stacey2018} found that both MG J0751+2716 and JVAS B1938+666 are undergoing a highly active star-forming phase, with SFRs of about 510 and 390~M$_{\odot}$ yr$^{-1}$, respectively (assuming a magnification for the FIR emission of 10 for JVAS B1938+666; see \citealt{Stacey2018} for discussion, and 16 for MG J0751+2716; \citealt{Alloin2007}). Therefore, our results are in favour for the presence of a coeval starburst with the central black hole growth \citep[e.g.][]{Walter2004}.

In order to understand the physical conditions of the star formation in starburst-quasar composites and the role of the AGN, spatially resolved observations of the material that fuels the star formation and AGN activity, namely the molecular gas, are needed. 
In only a few cases has the emission from the low $J$-level CO (2--1) molecular gas reservoirs of high redshift galaxies hosting an AGN been spatially resolved, and it was found to be distributed in rotating discs on scales of about 5~kpc in PSS J2322+1944 \citep{Riechers2008} and 15 kpc in RX J1131--1231 \citep{Paraficz2018}. For larger samples, where the CO (1--0) is not resolved, the sizes of the molecular gas reservoirs have been found to be $< 25$~kpc \citep{Riechers2011,Sharon2016}. On the other hand, in other starburst high-z galaxies hosting AGN, the molecular gas as traced by low and mid $J$-level CO lines revealed compact reservoirs  (kpc and sub-kpc scales) whose disturbed distribution and velocity fields suggested that the star formation could be triggered by gas-mergers or interactions \citep{Carilli2002, Walter2003, Carilli2007, Engel2010, Ivison2012}.

Our VLA observations, combined with the (modest) magnification from the gravitational lensing, are able to resolve the cold molecular gas traced by CO (1--0) in the case of JVAS B1938+666. We find that both lensed AGN have extended CO (1--0) emission with sizes $\geq 5$ kpc. Moreover, both molecular gas reservoirs are quite massive, with CO (1--0) derived masses of $M_{\rm gas}\geq 10^{10}$ M$_{\odot}$. These properties of the cold molecular gas, combined with the multi-wavelength morphology, suggest that both of the high redshift AGN are undergoing a phase of gas accretion, possibly due to an ongoing or recently completed major merger. The sizes of the molecular gas distribution are also comparable (in the case of JVAS~B1938+666) or slightly larger (MG J0751+2716) than those of the few well-studied unobscured high redshift quasars, in which it seems that the molecular gas is compact and at a higher excitation state \citep{Riechers2006, Weiss2007}. Furthermore, the gas distributions match the sizes of dust-obscured star-forming galaxies at high redshift, as for example, the bright dusty star-forming galaxy GN20 at $z=4.05$ \citep{Hodge2012}. Moreover, there is evidence of both lensed AGN having a deficiency of molecular gas emission in the proximity of the radio emission, and that the peak in the gas distribution is offset from the AGN and the host galaxy emission. This could imply that there is on-going star formation in those regions where the CO (1--0) has the maximum surface brightness, given the relation between the gas surface density and the SFR intensity, which would mean that the new stars are being formed in the disc and not in the central compact component detected at NIR wavelengths. We believe that our NIR and optical source reconstruction provides a good representation of the AGN host galaxy, since the radio emission in the case of MG~J0751+2716 detects the possible radio core, and hence the black hole at the centre, and for JVAS~B1938+666 we observe the two hotspots, whose mid-point is in the centre of the host galaxy. Furthermore, in both systems, the host galaxy has a fairly smooth surface brightness distribution. Finally, in the case of JVAS~B1938+666, the systemic velocity features are coincident with the centre of the host galaxy.

\begin{figure}
\centering
	\includegraphics[width = 0.5\textwidth]{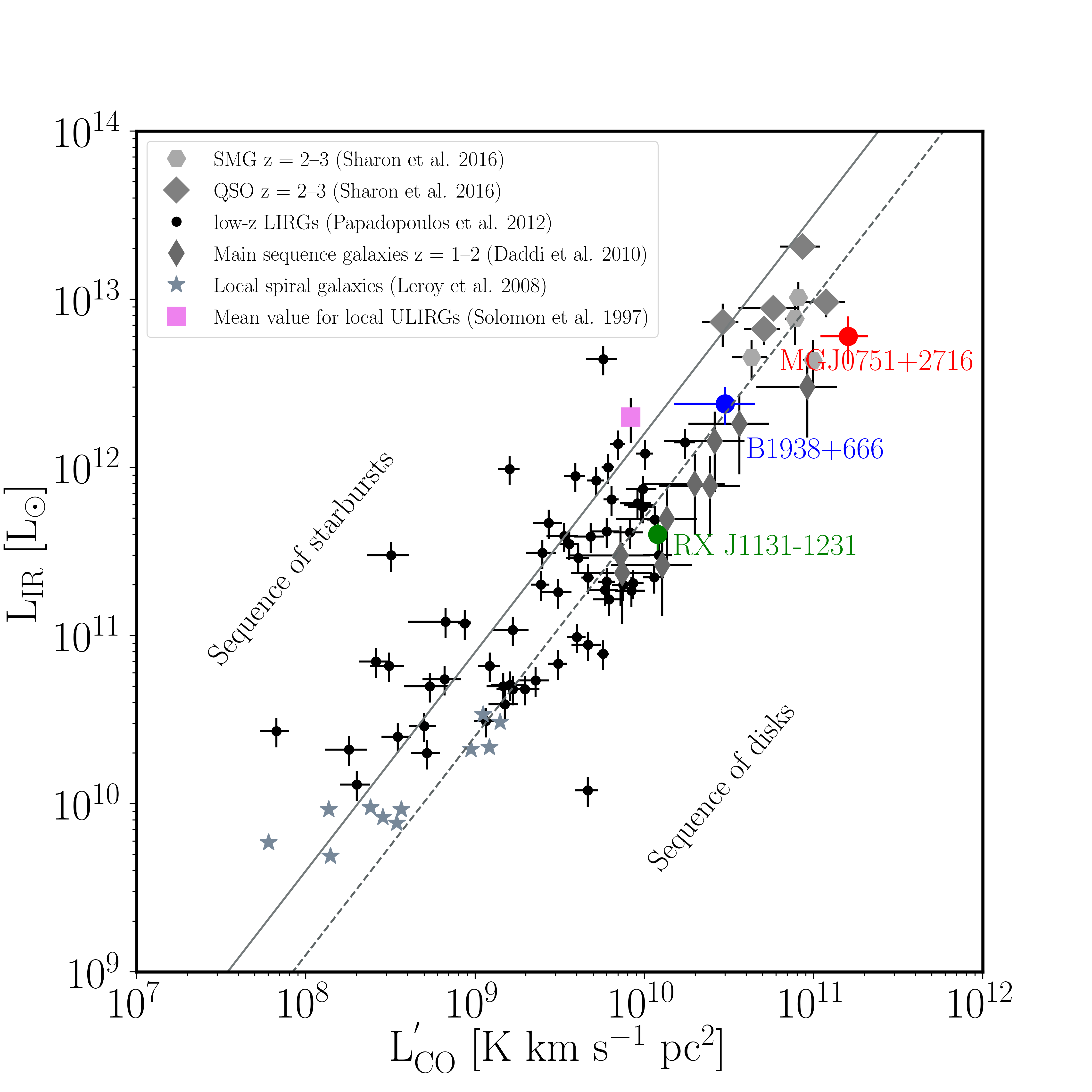}
    \caption{$L_{\rm IR}$ (integrated from 8 to 1000 $\mu$m) versus $L^\prime_{\rm CO}$ for different unlensed galaxy populations that have been detected in CO (1--0) and CO (2--1) line emission. The red and the blue filled circles are MG~J0751+2716 and JVAS~B1938+666, respectively. The green filled circle indicates the lensed disc-galaxy RX~J1131--1231 \citep{Paraficz2018}. The dashed line indicates the Kennicutt-Schmidt relation for local spiral galaxies, while the solid line represents the same relation for high-redshift disc galaxies \citep{Gao2004,Aravena2016}.} 
    \label{Fig:comparison}
\end{figure}

It has also been suggested that star formation has two distinct modes for disc and starburst galaxies, where spiral galaxies might have a long-lasting star formation while major mergers may be responsible for the rapid star formation process in ULIRGs (see Section \ref{Sec:alpha_CO}). In Fig.~\ref{Fig:comparison}, we show a variant of the Kennicutt-Schmidt relation ($L_{\rm IR}$--$L'_{\rm CO}$), which has the advantage of relying only on the observed quantities. We compare our two targets with dusty star-forming galaxies, main sequence galaxies, and those hosting an AGN (unobscured quasars). We find that JVAS B1938+666 and MG~J0751+2716 have $L'_{\rm CO}$ luminosities at the extreme end of the distribution that are higher than the main sequence galaxies, and that they are consistent with unobscured quasars and dusty star-forming galaxies at similar redshifts, however, this part of the parameter space has quite a large scatter and only a few measurements for each galaxy population. We also find that both lensed AGN have a lower IR luminosity with respect to the high redshift galaxy population, that is, they are forming fewer stars than expected given their abundance of molecular gas. Their position on the $L_{\rm IR}$--$L'_{\rm CO}$ plane may, therefore, imply that JVAS B1938+666 and MG J0751+2716 are less efficient at forming stars with respect to the population of unobscured quasars and dusty star-forming galaxies at comparable redshifts. This may be due to AGN feedback (as was also suggested for RX J1131$-$1231; \citealt{Paraficz2018}), or it could be due to the gas being replenished within the host galaxies, for example, via inflows or gas-rich mergers, with the large-scale star formation still to occur \citep[e.g.][]{Tacchella2016}. 

It is also possible that the sample of high redshift unobscured quasars have a significant contribution to the IR luminosity from AGN heating. Only high angular resolution imaging of the FIR dust emission will determine the location of the on-going star formation and any contribution from the AGN for our targets and the other objects shown in Fig.~\ref{Fig:comparison}. These observations will also determine robust magnifications for the dust emission for MG J0751+2716 and JVAS B1938+666. However, in the few cases imaged at mm-wavelengths so far, the heated dust emission is typically on the scale of 1-2 kpc, that is, a similar size to the stellar emission detected from the host galaxies investigated here.
Nevertheless, the dust emission in MG J0751+2716 and JVAS B1938+666 might be as extended as the CO (1--0) emission. If this is the case, our assumed FIR magnification factor would be consistent with $\mu_{\rm CO}$ for JVAS~B1938+666, while it would be eight times higher than $\mu_{\rm CO}$ for MG J0751+2716. This would lead to a much higher SFR derived by the FIR observations for MG J0751+2716. However, only FIR observations at high angular resolution can resolve the lensed images of the two systems, allowing us to properly take into account the differential magnification effect, which is particularly important in the infrared regime \citep{Serjeant2012}, and put more stringent constraints on the FIR emission in these two high redshift sources.

\subsection{Compact stellar cores at high redshift}

A consequence of the hierarchical galaxy formation process described above is that the inner part ($\sim 1$ kpc) of merger remnants should be dominated by a starburst component, which forms a central stellar cusp that is often interpreted as a precursor of the bulges/pseudo-bulges observed in the local Universe \citep{Kormendy1992, Hopkins2009b, Hopkins2013}. Also, gas-rich mergers with gas fractions larger than 30 per cent are necessary to produce such stellar density cores \citep{Robertson2006}. In addition, theoretical models predict that the majority of the classical bulges formed via gas-rich mergers should be already in place at redshifts 1 to 2 \citep{Conselice2007}. A way to test this formation scenario consists of identifying such compact stellar cores in gas-rich environments, which is challenging at high redshift. This difficulty is mainly due to the intrinsically low surface brightness of the cold molecular gas emission, coupled with the resolution needed to determine the structure of the rest-frame UV/optical emission from galaxies at $z>2$.  

The reconstructed host galaxies of the two lensed AGN studied here show evidence for compact rest-frame optical emission on scales of a few hundred pc (see Figs.~\ref{Fig:source_0751} and \ref{Fig:source_1938}) and indicative high gas-fractions larger than 50 per cent (see Fig.~\ref{Fig:gas_frac}).
In the case of MG J0751+2716, the S\'ersic index of the reddest component, which hosts the AGN, is close to a De Vaucouleurs profile, which may indicate that it already consists of a bulge, while the optical component of JVAS B1938+666 has a S\'ersic index of $1.5\pm 0.2$, which is more indicative of a pseudo-bulge/disc structure. According to this interpretation, these systems could represent two of the few high-redshift galaxies with possible resolved bulges and proto-disc structures \citep[e.g.][]{Labbe2003}, implying that at redshifts around 2 to 3, galactic bulges might be already formed. The presence of extended molecular gas reservoirs (discussed above) favours the inside-out formation for galactic bulges, namely that the bulge should form before the molecular gas disc \citep{ Conselice2007,Oldham2017, Tacchella2018}. However, in order to fully trace the ongoing star formation in MG~J0751+2716 and JVAS~B1938+666, it is fundamental to spatially resolve the heated dust for these two objects.

\section{Conclusions}
\label{sec:conclusions} 

In this paper, we presented high angular resolution observations with the VLA of the CO (1--0) molecular gas emission from the two gravitationally lensed star-forming/AGN composite galaxies MG J0751+2716 and JVAS B1938+666 at redshift 3.200 and 2.059, respectively. The detection from JVAS B1938+666, with a beam size of $0.25\times0.20$~arcsec$^2$ is currently the highest angular resolution detection of CO (1--0) from a high redshift object. We complement these data with radio continuum VLA, optical and NIR \textsl{HST} and NIR Keck-II adaptive optics observations, which are necessary to investigate the source properties, after correcting for the distortion due to gravitational lensing. 

From our multi-wavelength analysis, we find that both MG J0751+2716 and JVAS B1938+666 are heavily dust-obscured radio-loud AGN, with evidence of extreme star formation at the level of 510 and 390 M$_{\odot}$~yr$^{-1}$. The molecular gas as traced by the CO (1--0) emission has an elongated morphology with a size of $5\pm2$ kpc and $20\pm3$ kpc in JVAS B1938+666 and MG J0751+2716, respectively. The intrinsic velocity fields show distinct structures that can be associated with discs (elongated velocity gradients) and possible interacting objects (off-axis velocity components). We find intrinsic CO (1--0) luminosities of $L'_{\rm CO} = (2.5 \pm 0.8) \times 10^{10}$ K km s$^{-1}$ pc$^2$ and  $L'_{\rm CO} = (1.6\pm0.6) \times 10^{11}$ K km s$^{-1}$ pc$^2$ for JVAS B1938+666 and MG J0751+2716, respectively. These values are consistent with what has been observed in other star-forming dust-obscured active galaxies at similar redshifts, as for example, from the gravitationally lensed radio-quiet quasar APM 08279+5255 and the starburst galaxy GN20 \citep{Riechers2008, Hodge2012}.  

We estimate individual CO--H$_2$ conversion factors of $\alpha_{\rm CO} = 1.5 \pm 0.5$~(K km\,s$^{-1}$ pc$^2$)$^{-1}$ and $\alpha_{\rm CO} = 1.4 \pm 0.3$~(K km\,s$^{-1}$ pc$^2$)$^{-1}$, yielding molecular gas masses of $M_{\rm gas} = (2.5 \pm 0.8) \times 10^{11}$ M$_{\odot}$ and $M_{\rm gas} = (3.4 \pm 0.8) \times 10^{10 }$ M$_{\odot}$ for MG J0751+2716 and JVAS B1938+666, respectively.  Moreover, from our estimate of the dynamical masses, we infer gas fractions of about 60 per cent, which confirm the presence of a significant molecular gas reservoir.

Both of these two radio-selected objects show a compact central stellar component and small AGN radio jets on $\sim2$~kpc scales that are embedded in the extended molecular gas reservoirs discussed above. There is evidence for a decrease in the surface brightness of the CO (1--0) emission in the region close to the AGN emission in both objects, and particularly in the case of JVAS B1938+666, which indicates that the molecular gas at low excitation is not as abundant in those regions closer to the AGN. This could be taken as evidence for radiative feedback from the AGN, but only high angular resolution observations and the radiative transfer modelling of a set of high $J$-level CO transitions can test this hypothesis. Also, MG J0751+2716 and JVAS B1938+666 seem to lie at low IR luminosities in the Kennicutt-Schmidt relation, implying that they are forming stars at a lower level than expected given their abundant gas reservoirs. The reason for this is not clear, but may indicate the presence of AGN feedback that impedes the gas to form stars efficiently at the centre of the host galaxies. The compact stellar core of both AGN host galaxies, emitting at UV/optical rest-frame wavelengths, may be precursors of the central optical bulges/pseudo-bulges observed at present day.

The growth of galaxies is thought to be dominated by gas accretion processes. However, the cold molecular gas is still observationally difficult to access at high redshift. To date, the majority of the cold molecular gas detections are spatially unresolved or only marginally resolved at redshifts $>2$. Often high $J$-level transitions of CO are used to study the molecular gas content at high redshift galaxies, since they are brighter than the CO (1--0) emission, but they only probe the molecular gas at higher excitation and may not be representative of the total molecular gas content of these galaxies, which may lead to an underestimate of the dynamical masses, gas fractions and star formation properties. However, ALMA currently does not reach the crucial low $J$-level transitions of  CO for galaxies when the star formation and AGN activity peaked, limiting the study of the cold molecular gas at high redshift to a handful of bright objects. Moreover, spatially resolved observations of the cold molecular gas traced by the CO (1--0) emission on $<500$~pc-scales at redshifts $>2$ require a significant amount of observing time. Therefore, currently they can mainly be achieved by combining high sensitivity spectral line imaging and the magnifying power of gravitational lensing. Further observations, at similar angular resolutions to those presented here, for the sample of gravitationally lensed quasars with CO (1--0) detections, will establish whether the molecular gas associated with their host galaxies is extended, disc-like and abundant.

\section*{Acknowledgements}
We thank the anonymous referee for their useful and constructive comments on the paper.
CS is grateful for support from the National Research Council of Science and Technology, Korea (EU-16-001).  CS and JPM acknowledge support from the Netherlands Organization for Scientific Research (NWO, project number 629.001.023) and the Chinese Academy of Sciences (CAS, project number 114A11KYSB20170054).
SV has received funding from the European Research Council (ERC) under the European Union's Horizon 2020 research and innovation programme (grant agreement No 758853).
MWA acknowledges support from the Kavli Foundation.
LVEK is supported through an NWO-VICI grant (project number 639.043.308). 
CDF acknowledges support for this work from the National Science Foundation under Grant No. AST-1715611
DJL acknowledges support from the European Research Council (ERC) starting grant 336736- CALENDS. 
The data presented herein were obtained at the W. M. Keck Observatory, which is operated as a scientific partnership among the California Institute of Technology, the University of California and the National Aeronautics and Space Administration. The Observatory was made possible by the generous financial support of the W. M. Keck Foundation. The authors wish to recognize and acknowledge the very significant cultural role and reverence that the summit of Maunakea has always had within the indigenous Hawaiian community.  We are most fortunate to have the opportunity to conduct observations from this mountain.
The National Radio Astronomy Observatory is a facility of the National Science Foundation operated under cooperative agreement by Associated Universities, Inc. Based on observations made with the NASA/ESA Hubble Space Telescope, obtained from the Data Archive at the Space Telescope Science Institute, which is operated by the Association of Universities for Research in Astronomy, Inc., under NASA contract NAS 5-26555. These observations are associated with programs 7255, 7495 and 8268.



\bibliographystyle{mnras}
\bibliography{references} 



\appendix
\section{VLA visibility data}

\begin{figure*}
\centering
	\includegraphics[width =1.0\textwidth]{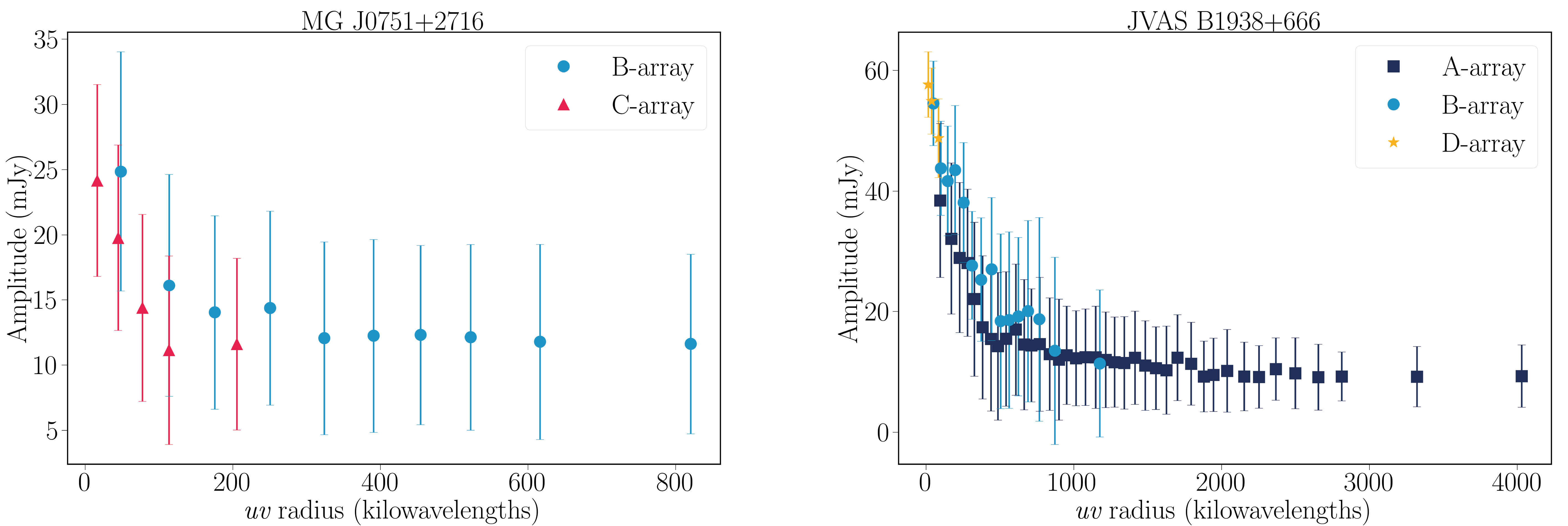}
    \caption{(Left) Visibility amplitude (mJy) as a function of radius in the {\it uv}-plane (projected baseline in kilowavelengths units) for the CO(1--0) and continuum emission in MG~J0751+2716. The bins have equal number of visibilities and the error bars plotted correspond to the standard deviation of each bin. We show the visibilities in the channels associated with the CO (1--0) emission as observed in B- (light blue circles) and C-configurations (red triangles). (Right) Same plot for the observed visibilities of JVAS~B1938+666 in A- (dark blue squares), B- (light blue circles) and D-configurations (yellow stars).} 
    \label{Fig:appendix}
\end{figure*}

The CO (1--0) emission has limited spatial information for both systems due to the VLA resolving out the emission with the longest baselines. This can be seen as a rapid decrease in flux density with increasing $uv$ distance, as illustrated in Fig~\ref{Fig:appendix}.

\bsp	
\label{lastpage}
\end{document}